\documentclass[11pt,letterpaper]{article}
\usepackage{jheppub}
\usepackage{amsmath}
\usepackage{graphicx}
\usepackage{amssymb}
\usepackage{color}      

\newcommand{\beq}{\begin{equation}}
\newcommand{\eeq}{\end{equation}}
\newcommand{\bea}{\begin{eqnarray}}
\newcommand{\eea}{\end{eqnarray}}
\newcommand{\nn}{\nonumber}

\preprint{
{\vbox {
\hbox{\bf MSUHEP-19-004}
}}}
\vspace*{0.2cm}

\title{Soft Gluon Resummation in $t$-channel  single top quark production at the LHC}

\author{Qing-Hong Cao$^{1,2,3}$, Peng Sun$^{4}$, Bin Yan$^{5}$, C.-P. Yuan$^5$, Feng Yuan$^6$}

\affiliation{$^1$Department of Physics and State Key Laboratory of Nuclear Physics and Technology, Peking University, Beijing 100871, China}
\affiliation{$^2$Collaborative Innovation Center of Quantum Matter, Beijing 100871, China}
\affiliation{$^3$Center for High Energy Physics, Peking University, Beijing 100871, China}
\affiliation{$^4$Department of Physics and Institute of Theoretical Physics, Nanjing Normal University, Nanjing, Jiangsu, 210023, China}
\affiliation{$^5$Department of Physics and Astronomy,
Michigan State University, East Lansing, MI 48824, USA}
\affiliation{$^6$Nuclear Science Division, Lawrence Berkeley National
Laboratory, Berkeley, CA 94720, USA}

\emailAdd{qinghongcao@pku.edu.cn}
\emailAdd{pengsun@msu.edu}
\emailAdd{yanbin1@msu.edu}
\emailAdd{yuan@pa.msu.edu}
\emailAdd{fyuan@lbl.gov}

\abstract{
We present a detailed phenomenological study of the multiple soft gluon radiation
for the $t$-channel single top  and anti-top quark production at the Large Hadron Collider (LHC).
By applying the transverse momentum dependent factorization formalism,
large logarithms introduced by small total transverse momentum
$q_\perp$  of the single-top (anti-top) plus one-jet final state system are resummed
to all orders in the expansion of the strong interaction coupling at
the accuracy of Next-to-Leading Logarithm.
We discuss various kinematical distributions which are sensitive to this effect and find that 
soft gluon radiation become  more important when the final state jet is required to be
in the forward region.  We show that the main difference from PYTHIA prediction lies on
the inclusion of the exact color coherence effect between the
initial and final states in our resummation calculation.
We further propose to apply the  experimental observable $\phi^*$ to test the effect of multiple gluon
radiation in the single-top and anti-top events. The bottom quark mass effect and jet rapidity distribution are also discussed.

}


\begin{document}

\maketitle

\section{Introduction}
Top quark has its mass around the electroweak symmetry breaking scale and is expected to play an important role of testing the Standard Model (SM) and provide a window to  new physics (NP) of beyond the SM.  At the CERN Large Hadron Collider (LHC), top quarks can be produced by the electroweak gauge interaction~\cite{Dawson:1984gx,Willenbrock:1986cr,Dawson:1986tc,Yuan:1989tc}. Its cross section is proportional to the $Wtb$ coupling~\cite{Kane:1991bg,Carlson:1994bg,Chen:2005vr,Cao:2007ea,Berger:2009hi,Fabbrichesi:2014wva,Bernardo:2014vha,Cao:2015doa,Prasath:2014mfa,Hioki:2015env,Zhang:2016omx,
Birman:2016jhg,Boos:2016zmp,Jueid:2018wnj}  and Cabibbo-Kobayashi-Maskawa (CKM) matrix element $V_{tb}$ ~\cite{Berger:2009hi,Cao:2015qta}. Hence, it offers a promising way to  constrain  various $Wtb$ anomalous couplings, induced by NP, or   determine $V_{tb}$ without assuming  the generation of  quarks. Furthermore, top quarks produced from the electroweak process are highly polarized, and  the degree of  polarization of the top quark in the single top quark process events can be used to discriminate New Physics models at the LHC~\cite{Ladinsky:1992vv,Carlson:1993dt,Carlson:1995ck,Tait:1996dv,Li:1996ir,Li:1997qf,Tait:1997fe,He:1998ie,Tait:2000sh,Malkawi:1996fs,Hsieh:2010zr,Cao:2012ng,Drueke:2014pla,Cao:2013ud,Cao:2006wk,Berger:2011hn,Berger:2011xk}, e.g. the $W^\prime$ gauge boson from $G(221)$ models~\cite{Malkawi:1996fs,Hsieh:2010zr,Cao:2012ng}, new heavy
fermions~\cite{Cao:2006wk,Berger:2011hn,Berger:2011xk} and charged scalars~\cite{Drueke:2014pla,Cao:2013ud}.
In addition, single top quark production can also be used to measure  top quark mass~\cite{Alekhin:2016jjz,CMS-PAS-TOP-15-001} and constrain the light quark PDFs~\cite{Alekhin:2015cza}. Therefore, a precise study of both the inclusive and differential cross sections of single top quark events is vital to test the SM and search for  NP.

At the LHC, single top quark is predominantly produced through $t$-channel mode. The total and differential cross sections have been measured by the ATLAS and CMS collaborations at $\sqrt{s}=7~{\rm TeV}$~\cite{Aad:2014fwa,Chatrchyan:2012ep}, $\sqrt{s}=8~{\rm TeV}$~\cite{Aaboud:2017pdi,Khachatryan:2014iya} and $\sqrt{s}=13~{\rm TeV}$~\cite{Aaboud:2016ymp,Sirunyan:2016cdg}.  These results are consistent with the SM predictions~\cite{Brucherseifer:2014ama,Berger:2016oht,Berger:2017zof}. The single top quark production and decay in hadron collision
at the next-to-leading order (NLO) accuracy in QCD correction,
has been known for many years
~\cite{Zhu:2001hw,Harris:2002md,Cao:2004ky, Cao:2004ap,Campbell:2004ch,Cao:2005pq, Campbell:2005bb,Cao:2008af,Heim:2009ku,Campbell:2009ss,Schwienhorst:2010je,Falgari:2010sf, Frixione:2005vw,Alioli:2009je,Frederix:2012dh,Frederix:2016rdc,Carrazza:2018mix,deBeurs:2018pvs}. Recently, the dominant part of the next-to-next-to-leading order (NNLO) QCD correction to
predicting the detailed kinematical distributions, including proper spin correlations, in
$t$-channel single top events, has been
discussed in Refs.~\cite{Brucherseifer:2014ama,Berger:2016oht,Berger:2017zof}. Beyond the fixed order calculation,
the threshold resummation technique is also widely discussed to imporve the inclusive production rate of the single-top quark event in the literatures~\cite{Kidonakis:2006bu,Kidonakis:2007ej,Zhu:2010mr,Wang:2010ue,Kidonakis:2011wy,Wang:2012dc}. Furthermore, the accuracy of  transverse momentum distribution of the top quark could also be improved by summing over  large logarithms $\ln(m_t^2/s_4)$ with $s_4\to 0$, where $s_4=\hat{s}+\hat{t}+\hat{u}-m_t^2$, in which  $\hat{s}$, $\hat{t}$ and $\hat{u}$  are the usual Mandelstam variables~\cite{Kidonakis:2006bu,Kidonakis:2007ej,Zhu:2010mr,Wang:2010ue,Kidonakis:2011wy,Wang:2012dc}. 

In a recent publication~\cite{Cao:2018ntd},  we studied  kinematical distributions
of $t$-channel single top events by applying  the transverse momentum  ($q_\perp$)  resummation  formalism to sum over
large logarithms $\ln(Q^2/q_{\perp}^2)$, with $Q \gg q_{\perp}$,
to all orders in the expansion of the strong interaction coupling ($g_s$) at
the next-to-leading-logarithm (NLL) accuracy. Here, $Q$ and $q_\perp$ are the
invariant mass $Q$ and total transverse momentum
$q_\perp$  of the single-top plus one-jet final state system, respectively.
The $q_\perp$ resummation technique is based on the transverse momentum dependent (TMD)
factorization formalism~\cite{Collins:1984kg}, which has been widely discussed in the color singlet processes, 
such as Drell-Yan  production~\cite{Collins:1981uk,Collins:1981va}.
Extending the $q_\perp$ resummation formalism to processes with more complex 
color structure are also discussed widely recently, e.g. heavy quark production~\cite{Zhu:2012ts,Li:2013mia,Zhu:2013yxa}, and processes involving multijets in the final state~\cite{Sun:2014lna,Sun:2014gfa,Sun:2015doa,Sun:2016mas,Sun:2016kkh,Xiao:2018esv,Cao:2018ntd,Sun:2018beb,Sun:2018icb,Sun:2018byn,Buffing:2018ggv,Liu:2018trl}. For those processes,  additional large logarithms,  induced by  soft gluon radiation from  color coherence effects, need to be resummed via the modified $q_\perp$ resummation formalism.

As  shown on the Ref~\cite{Cao:2018ntd}, the $q_\perp$ spectrum of $t$-channel single top quark production strongly depends on the color coherence effects between the initial state and final state jets and the treatment of bottom quark mass in the resummation calculation.
The sub-leading logarithms from soft gluon interaction play an even more important role
when the final state jet is required to be in the forward region, which is identified as the
$t$-channel single top signal region,
where our resummation prediction is different with PYTHIA parton shower result.
In the present paper, we present details of our results in Ref.~\cite{Cao:2018ntd}. In Sec. 2, we present the detail of our resummation calculation. Its phenomenology is discussed in Sec. 3. Our conclusion is given in Sec. 4. 

\section{Factorization and Resummation}
We consider the process $pp\to t(\bar{t}) +jet+X$ at the LHC.
Using the TMD resummation formalism  presented
in Ref.~\cite{Sun:2015doa}, the differential cross section of the
$t$-channel single top quark production process
can be written as
\begin{align}
\frac{d^4\sigma}
{dy_t dy_J d P_{J\perp}^2
d^2q_{\perp}}=\sum_{ab}
\left[\int\frac{d^2\vec{b}}{(2\pi)^2}
e^{-i\vec{q}_\perp\cdot
\vec{b}}W_{ab\to t J}(x_1,x_2,\textbf{b})+Y_{ab\to tJ}\right] \ ,\label{resumy}
\end{align}
where $y_t$ and $y_J$ are the rapidities of  the top quark and the final state jet, respectively;
 $P_{J\perp}$ and $q_{\perp}$ are the transverse momenta of the jet and the total transverse momentum of the top quark and the jet system, i.e. $\vec{q}_\perp=\vec{P}_{t\perp}+\vec{P}_{J\perp}$. The $W_{ab\to tJ}$ term contains all order
 resummation contribution, in powers of $\ln(Q^2/q_{\perp}^2)$, and the inclusion of
 the $Y_{ab\to tJ}$ term is to account for the part of fixed-order corrections which are not included in
 the expansion of the $W_{ab\to tJ}$ term to the same order in strong coupling constant $g_s$.
 The variables $x_1, x_2$ are momentum fractions of the incoming hadrons carried by the partons, with
\begin{equation}
x_{1,2}=\frac{\sqrt{m_t^2+P^2_{t\perp}}e^{\pm y_t}+\sqrt{P^2_{J\perp}}e^{\pm y_J}}{\sqrt{S}},
\end{equation}
where $m_t$ and $S$ are the top quark mass and squared collider energy, respectively.

The above $W$ term can be further written as
\begin{eqnarray}
W_{ab\to tJ}\left(x_1,x_2,\textbf{b}\right)&=&x_1\,f_a(x_1,\mu_F=b_0/b_*)
x_2\, f_b(x_2,\mu_F=b_0/b_*) e^{-S_{\rm Sud}(Q^2,\mu_{\rm Res},b_*)}e^{-\mathcal{F}_{NP}(Q^2,\textbf{b})} \nonumber\\
&\times& \textmd{Tr}\left[\mathbf{H}_{ab\to tJ}(\mu_{\rm Res})
\mathrm{exp}[-\int_{b_0/b_*}^{\mu_{\rm Res}}\frac{d
\mu}{\mu}\mathbf{\gamma}_{}^{s\dag}]\mathbf{S}_{ab\to tJ}(b_0/b_*)
\mathrm{exp}[-\int_{b_0/b_*}^{\mu_{\rm Res}}\frac{d
\mu}{\mu}\mathbf{\gamma}_{}^{s}]\right]\ ,\nn\\\label{resum}
\end{eqnarray}
where $Q^2=\hat{s}=x_1x_2S$, the hard scale of this process, $b_0=2e^{-\gamma_E}$  with $\gamma_E$ being the Euler constant 0.5772, 
$f_{a,b}(x,\mu_F)$ are parton distribution functions (PDF) for the incoming partons $a$ and $b$,
and $\mu_{\rm Res}$ represents the resummation scale of this process.
Here, $b_*=\textbf{b}/\sqrt{1+\textbf{b}^2/b_{\rm{max}}^2}$ with $b_{\rm {max}}=1.5~{\rm GeV}^{-1}$,
which is introduced to factor out the non-perturbative contribution
$e^{-\mathcal{F}_{NP}(Q^2,b)}$, arising from the large $\textbf{b}$ region (with $\textbf{b} \gg b_*$)
~\cite{Landry:1999an,Landry:2002ix,Sun:2012vc,Su:2014wpa},
\begin{equation}
\mathcal{F}_{NP}(Q^2,\textbf{b})=g_1\textbf{b}^2+g_2\ln\dfrac{Q}{Q_0}\ln\dfrac{\textbf{b}}{b_*},
\end{equation}
where $g_1=0.21$, $g_2=0.84$ and $Q_0^2=2.4~{\rm GeV}^2$~\cite{Su:2014wpa}.
In this study, we shall use CT14NNLO PDFs~\cite{Dulat:2015mca} for our numerical calculation.
Hence, our resummation calculation should be consistently done in the
General-Mass-Variable-Flavor (GMVR) scheme that the PDFs are
determined.
The bottom quark PDF is set to zero when the factorization scale $\mu_F$
is below the bottom quark mass threshold, i.e. $\mu_F< m_b$.
To properly describe the small $q_\perp$ region (for $q_\perp<m_b$), the S-ACOT scheme~\cite{Aivazis:1993kh,Aivazis:1993pi,Collins:1998rz,Kramer:2000hn} is
adopted to account for the effect from the (non-zero) mass of the incoming bottom quark
in the hard scattering process.
In Refs.~\cite{Nadolsky:2002jr,Belyaev:2005bs,Berge:2005rv}, a detailed discussion has been given on how to implement the
S-ACOT scheme in the $q_\perp$ resummation formalism, for processes initiated by
bottom quark fusion.
In short,
the S-ACOT scheme retains massless quark in the calculation of
the hard scattering amplitude (of $q b \to q^{\prime}t$),
 but with the (bottom quark) mass dependent Wilson coefficient $C_{b/g}^{(1)}(x,\textbf{b},\mu_F)$,
 to account for the contribution from gluon splitting into a $b\bar{b}$ pair~\cite{Nadolsky:2002jr,Belyaev:2005bs,Berge:2005rv},
\begin{align}
C_{b/g}^{(1)}(z,\textbf{b},m_b,\mu_F)=\dfrac{1}{2}z(1-z)\textbf{b}m_bK_1(\textbf{b}m_b)
+P_{q/g}^{(1)}(z)
\left[K_0(\textbf{b}m_b)-\theta(\mu_F-m_b)\ln\dfrac{\mu_F}{m_b}\right]
\label{eq:cf}
\end{align}
where the factorization scale $\mu_F=b_0/\textbf{b}$, $K_0(z)$ and $K_1(z)$ are the modified Bessel functions, $P_{q/g}^{(1)}(z)$ is the gluon splitting kernel. This expression reduces to the massless result when 
$1/\textbf{b}\ll m_b$~\cite{Belyaev:2005bs},
\begin{align}
&C_{b/g}^{(1)}(z,\textbf{b},m_b=0,\mu_F)=\dfrac{1}{2}z(1-z)-\ln\dfrac{\mu_F\textbf{b}}{b_0}P_{q/g}^{(1)}(z).
\label{eq:cf2}
\end{align}
The hard and soft factors $\mathbf{H}$ and $\mathbf{S}$ are expressed as matrices in the color space of  $ab\to tJ$, and $\gamma^s$ is the associated anomalous dimension of the soft factor. The Sudakov form factor ${\cal S}_{\rm Sud}$ resums the leading double logarithm and the sub-leading logarithms,
\begin{align}
S_{\rm Sud}(Q^2,\mu_{\rm Res},b_*)=\int^{\mu_{\rm Res}^2}_{b_0^2/b_*^2}\frac{d\mu^2}{\mu^2}
\left[\ln\left(\frac{Q^2}{\mu^2}\right)A+B 
+D_1\ln\frac{Q^2-m_t^2}{P_{J\perp}^2R^2}+
D_2\ln\frac{Q^2-m_t^2}{m_t^2}\right]\ , \label{su}
\end{align}
where $R$ represents the cone size of the final state jet. Here the parameters $A$, $B$, $D_1$ and $D_2$ can be expanded perturbatively in $\alpha_s$, which is $g_s^2/(4 \pi)$. At one-loop order,
\begin{equation}
A=C_F\dfrac{\alpha_s}{\pi},\quad B=-2C_F\dfrac{\alpha_s}{\pi},\quad D_1=D_2=C_F\dfrac{\alpha_s}{2\pi},
\end{equation}
with $C_F=4/3$.
In our numerical calculation, we will also include the $A^{(2)}$ contribution , because  it is associated with the incoming parton and is universe for all hard processes initialed by the same incoming parton~\cite{Catani:2000vq},
\bea
A^{(2)}=C_F\left(\dfrac{\alpha_s}{\pi}\right)^2\left[\left(\dfrac{67}{36}-\dfrac{\pi^2}{12}C_A\right)-\dfrac{5}{18}N_f\right],
\eea
where $C_A=3$ and  $N_f=5$ is the number of effective light quarks. The cone size $R$ is introduced to regulate the collinear gluon radiation associated with the final state jet~\cite{Sun:2014lna,Sun:2014gfa,Sun:2015doa,Sun:2016mas,Sun:2016kkh,Xiao:2018esv,Cao:2018ntd,Sun:2018beb}.

The soft gluon radiation can be factorized out based on the Eikonal approximation method, i.e. for each incoming and outgoing color particles, the soft gluon radiation is factorized into an associated gauge link along the particle momentum direction. The color correlation between the color particles in this process is expanded by a group of orthogonal color bases.
For the $t$-channel single top quark production, there are two orthogonal color configurations,
\begin{equation}
C_{1kl}^{ij}=\delta_{ik}\delta_{jl},\quad C_{2kl}^{ij}=T^{a^\prime}_{ik}T^{a^\prime}_{jl},
\label{eq:basis}
\end{equation}
where $i,j$ are color indices of the two incoming partons, $k,l$ are color indices of the jet and the top quark in final states and $a^\prime$ is color index of the gluon. We follow the procedure of Ref.~\cite{Sun:2015doa} to calculate the soft factor. Its definition in such color basis can be written as,
\begin{align}
S_{IJ}&=\int_0^\pi
\frac{d\phi}{\pi}\; C^{bb'}_{Iii'} C^{aa'}_{Jll'}\langle 0|{\cal L}_{
vcb'}^\dagger(b) {\cal
L}_{\bar vbc'} (b){\cal L}_{\bar
vc'a'}^\dagger(0)
 {\cal
L}_{ vac}(0) {\cal L}_{n ji}^\dagger(b) {\cal
L}_{\bar n i'k}(b) {\cal L}_{\bar nkl}^\dagger (0) {\cal
L}_{nl'j} (0)  |0\rangle \ ,\label{soft}
\end{align}
where we integrated out the azimuthal angle of the top quark and traded the relative azimuthal angle $\phi$ for  the $q_\perp$. $I$ and $J$ represent the color basis index, $n$ and $\bar{n}$ represent  the momentum directions of the top quark and the jet in this process,  $v$ and $\bar{v}$ are the momentum directions of the initial states. The gauge link ${\cal L}_{v}(\xi)$ is defined along the $v$ direction
\begin{equation}
{\cal L}_{v}(\xi)\equiv P exp\left(-ig_s\int_{-\infty}^0d\lambda v\cdot A(\lambda v+\xi)\right).
\end{equation}
In this basis, the LO soft function is given by
\begin{eqnarray}
\mathbf{S}^{(0)}=  \left[
             \begin{array}{cc}
               C_A^2 & 0 \\
               0 & \frac{C_AC_F}{2} \\
             \end{array}
           \right] \ ,
\end{eqnarray}
At the NLO level, the soft function can be represented as
\begin{equation}
\mathbf{S}^{(1)}=\sum_{i,j}\mathbf{W}_{ij}I_{ij},
\end{equation}
where $\mathbf{W}_{ij}$ is the color matrix element. For subprocess $ub\to dt$~\cite{Sun:2015doa}
\begin{align}
W_{11}=&W_{22}=C_F\mathbf{S}^{(0)},\nn\\
W_{12}=&W_{34}=\left[
             \begin{array}{cc}
                0                              &- \dfrac{C_AC_F}{2} \\
                -\dfrac{C_AC_F}{2}  & \dfrac{C_F}{2} \\
             \end{array}
           \right] \,\nn\\
W_{13}=&W_{24}=\left[
             \begin{array}{cc}
                C_A^2C_F                & 0\\
               0                               & -\dfrac{C_F}{4} \\
             \end{array}
           \right] \ ,\nn\\  
W_{14}=&W_{23}=\left[
             \begin{array}{cc}
                0                              & \dfrac{C_AC_F}{2} \\
                \dfrac{C_AC_F}{2}  & \dfrac{1}{4}(C_A^2-2)C_F \\
             \end{array}
           \right] \ .    
\end{align}
For subprocess $\bar{d} b\to \bar{u}t$, 
\begin{align}
W_{11}=&W_{22}=C_F\mathbf{S}^{(0)},\nn\\
W_{12}=&W_{34}=\left[
             \begin{array}{cc}
                0                              & \dfrac{C_AC_F}{2} \\
                \dfrac{C_AC_F}{2}  & \dfrac{1}{4}(C_A^2-2)C_F \\
             \end{array}
           \right] \ ,\nn\\
W_{13}=&W_{24}=\left[
             \begin{array}{cc}
                C_A^2C_F                & 0\\
               0                               & -\dfrac{C_F}{4} \\
             \end{array}
           \right] \ ,\nn\\  
W_{14}=&W_{23}=\left[
             \begin{array}{cc}
                0                              &- \dfrac{C_AC_F}{2} \\
                -\dfrac{C_AC_F}{2}  & \dfrac{C_F}{2} \\
             \end{array}
           \right]\  .                  
\end{align}
$I_{ij}$ represents the kinematic integral for the soft gluon radiation between $i$ and $j$ gauge links,
\begin{eqnarray}
I_{ij}&=&\dfrac{\alpha_s}{2\pi^2}\int_0^{\pi}\dfrac{(\sin\phi)^{-2\epsilon}d\phi}{\frac{\sqrt{\pi}\Gamma(1/2-\epsilon)}{\Gamma(1-\epsilon)}}\int dk^+dk^-\dfrac{n_i\cdot n_j}{(k\cdot n_i)(k\cdot n_j)}\delta(k^2)\theta(k_0),
\end{eqnarray}
where $k$ represents the radiated gluon momentum, $n_{i,j}$  are dimensionless vectors along the directions of momentum $p_{i,j}$.
The soft factor of the $t$-channel single top quark production process at the NLO level is
\begin{align}
\mathbf{S}^{(1)}=& -\frac{\alpha_s}{2\pi}\mathbf{S}^{(0)}\;C_F\ln\dfrac{\mu^2b_*^2}{b_0^2}
\left[-1
+\ln\left(\frac{\hat{s}-m_t^2}{R^2P_{J\perp}^2}\frac{\hat{s}-m_t^2}{m_t^2}\right)\right] 
-\frac{\alpha_s}{2\pi}\left[2\;\Xi\ln\dfrac{\mu^2b_*^2}{b_0^2}+S^{\epsilon}\right],
\label{sud12}
\end{align}
where $\Xi$ is process dependent, with
\begin{align}
\Xi_{ub\to dt}&=\dfrac{1}{2}\left[\begin{array}{cc}
      2C_FC_A^2\,T &\;\; C_FC_A\,U \\\\
       C_FC_A\,U &\;\; \frac{1}{2}(C_A^2-2)C_F\,U-\frac{C_F}{2}\,T \\
     \end{array} \right ]\ &,\\
\Xi_{\bar{d}b\to\bar{u}t}&=\dfrac{1}{2}\left[
     \begin{array}{cc}
       2C_FC_A^2\,T &\;\; -C_FC_A\,U \\\\
       -C_FC_A\,U &\;\; C_F\,U-\frac{C_F}{2}\,T \\
     \end{array}
   \right]\ ,
\end{align}
and
\begin{align}
T&=\ln(\dfrac{-\hat{t}}{\hat{s}})+\ln(\dfrac{-(\hat{t}-m_t^2)}{\hat{s}-m_t^2}),\quad
U=\ln(\dfrac{-\hat{u}}{\hat{s}})+\ln(\dfrac{-(\hat{u}-m_t^2)}{\hat{s}-m_t^2}).
\label{sud15}
\end{align}
Here $\hat{s}=x_1x_2S$, $\hat{t}=(p_u-p_d)^2$, $\hat{u}=(p_b-p_d)^2$ for the $ub\to dt$ process, while for the $\bar{d}b\to \bar{u} t$ process, $\hat{t}=(p_{\bar{d}}-p_{\bar{u}})^2$ and $\hat{u}=(p_b-p_{\bar{u}})^2$.
The matrix $S^{\epsilon}$ is also process dependent,
\begin{align}
S^{\epsilon}_{ub\to dt}&=C_F\left[\begin{array}{cc}
      C_A^2 (I_S+I_L) &\;\; -(I_{3,4}-I_S)\dfrac{C_A}{2} \\
      -(I_{3,4}-I_S)\dfrac{C_A}{2} &\;\; \dfrac{1}{2}I_{3,4}+\dfrac{(C_A^2-3)}{4}I_S+\dfrac{C_AC_F}{2}I_L \\
     \end{array} \right ], \,\\
S^{\epsilon}_{\bar{d}b\to\bar{u}t}&=C_F\left[\begin{array}{cc}
      C_A^2 (I_S+I_L) &\;\; (I_{3,4}-I_S)\dfrac{C_A}{2} \\
      (I_{3,4}-I_S)\dfrac{C_A}{2} &\;\; \dfrac{(C_A^2-2)}{4}I_{3,4}+\dfrac{1}{4}I_S+\dfrac{C_AC_F}{2}I_L \\
     \end{array} \right ], \
\end{align}
where
\begin{equation}
I_L=\ln\dfrac{m_t^2}{m_t^2+P_{J\perp}^2},\quad
I_S=\left[\dfrac{1}{2}\ln^2(\dfrac{1}{R^2})-\mathrm{Li}_2(\dfrac{-P_{J\perp}^2}{m_t^2})\right].
\end{equation}
The $I_{3,4}$ represents the kinematic integral for the soft gluon radiation between the
final state jet and top quark gauge links~\cite{Sun:2018byn}, and 
\begin{align}
I_{3,4}&=-\mathrm{Li}_2\dfrac{m_t^2+\hat{t}-\hat{u}}{\hat{t}}-\mathrm{Li}_2\dfrac{(2m_t^2-\hat{s})(m_t^2-\hat{t})}{\hat{s}\hat{t}}+\mathrm{Li}_2\dfrac{(\hat{s}-2m_t^2)\hat{t}}{\hat{s}\hat{u}}-\ln\dfrac{m_t^2-\hat{u}}{m_t^2+\hat{t}-\hat{u}}\ln\dfrac{-m_t^2(m_t^2+\hat{t}-\hat{u})}{\hat{s}\hat{u}}\nn\\
&+\ln\dfrac{-\hat{t}}{m_t^2+\hat{t}-\hat{u}}\ln\dfrac{(m_t^2-\hat{s})(m_t^2+\hat{t}-\hat{u})}{\hat{s}\hat{u}}
+(\hat{t}\leftrightarrow\hat{u})-\ln\dfrac{\hat{s}-m_t^2}{m_t^2}\ln\dfrac{\hat{t}\hat{u}}{m_t^4-(\hat{t}-\hat{u})^2}\nn\\
&-\ln\dfrac{P_{J\perp}^2R^2\hat{s}}{\hat{t}\hat{u}}\ln\dfrac{\hat{s}-m_t^2}{-P_{J\perp}^2R^2}-\dfrac{1}{2}\ln^2\dfrac{P_{J\perp}^2R^2}{\hat{s}-2m_t^2}
-\dfrac{1}{2}\ln^2\dfrac{m_t^2}{2m_t^2-\hat{s}}+\dfrac{1}{2}\ln^2\dfrac{\hat{s}-m_t^2}{2m_t^2-\hat{s}}-\ln\dfrac{\hat{s}-m_t^2}{2m_t^2-\hat{s}}\ln\dfrac{P_{J\perp}^2R^2}{\hat{s}-2m_t^2}\nn\\
&+2\ln\dfrac{\hat{
s}-m_t^2}{2m_t^2-\hat{s}}\ln\dfrac{P_{J\perp}^2R^2}{\hat{s}-m_t^2}+\ln\dfrac{m_t^2}{2m_t^2-\hat{s}}\ln\dfrac{m_t^2\hat{s}}{\hat{t}\hat{u}}-2\ln\dfrac{2m_t^2-\hat{s}}{m_t^2-\hat{s}}\ln\dfrac{m_t^2}{2m_t^2-\hat{s}}-2\mathrm{Li}_2\dfrac{m_t^2}{\hat{s}-m_t^2}-\dfrac{\pi^2}{3}.
\end{align}
In the massless limit of the top quark, it will recover the result of dijet production, as  in Ref.~\cite{Sun:2015doa},
\begin{equation}
I_{3,4}=\dfrac{1}{2}\ln^2\dfrac{1}{R_1^2}+\dfrac{1}{2}\ln^2\dfrac{1}{R_2^2}+\dfrac{\pi^2}{3}-4\ln\dfrac{-\hat{s}}{\hat{t}}\ln\dfrac{-\hat{s}}{\hat{u}},
\end{equation}
where $R_1$ and $R_2$ are the radii of the two cone jets. 

$S_{IJ}$ satisfies the renormalization group equation,
\begin{equation}
\dfrac{d}{d\ln\mu}S_{IJ}(\mu)=-\sum_LS_{IL}\Gamma^S_{LJ}-\sum_L\Gamma^{S\dagger}_{IL}S_{LJ},
\end{equation}
where
\begin{equation}
\mathbf{\Gamma^S}=\dfrac{\alpha_s}{2\pi}\left[-1+\ln\dfrac{\hat{s}-m_t^2}{R^2P_{J\perp}^2}+\ln\dfrac{\hat{s}-m_t^2}{m_t^2}\right]C_F\mathbf{\Gamma^E}+\mathbf{\gamma^S}.
\end{equation}
Here $\mathbf{\Gamma^E}$ is an identity matrix and $\mathbf{\gamma^S}$ is  the associated anomalous dimension.
We obtain
\begin{align}
\mathbf{\gamma^S}_{ub\to dt}&=\dfrac{\alpha_s}{\pi}\left[\begin{array}{cc}
      C_F\,T &\;\; C_F/C_A\,U \\\\
       U &\;\; \frac{1}{2}(C_A-2/C_A)\,U-\frac{1}{2C_A}\,T \\
     \end{array} \right ]\ ,\nn\\
\mathbf{\gamma^S}_{\bar{d}b\to\bar{u}t}&=\dfrac{\alpha_s}{\pi}\left[
     \begin{array}{cc}
       C_F\,T &\;\; -\dfrac{C_F}{2C_A}\,U \\\\
       -U &\;\; -\frac{1}{2C_A}\,(T-2U) \\
     \end{array}
   \right]\ .
\label{eq:gammas}
\end{align}

The jet function originated from  the collinear gluon radiation of jet . In this work, we apply the anti-$k_T$ jet algorithm as in Refs.~\cite{Mukherjee:2012uz,Sun:2015doa}, and 
\begin{equation}
J_q=\dfrac{\alpha_sC_F}{2\pi \Gamma(1-\epsilon)}\left[\dfrac{1}{\epsilon^2}+\dfrac{1}{\epsilon}\left(\dfrac{3}{2}-\ln\dfrac{P_{J\perp}^2R^2}{\mu_{\rm Res}^2}\right)+I_q\right],
\label{eq:Jet}
\end{equation}
where $\epsilon = (D - 4)/2$ in D-dimensional regularization, and the finite term $I_q$ is
\begin{equation}
I_q=\dfrac{1}{2}\left(\ln\dfrac{P_{J\perp}^2R^2}{\mu_{\rm Res}^2}\right)^2-\dfrac{3}{2}\ln\dfrac{P_{J\perp}^2R^2}{\mu_{\rm Res}^2}+\dfrac{13}{2}-\dfrac{2}{3}\pi^2.
\end{equation}
The singular terms in the jet function are independent  of  jet algorithm, while
the finite term $I_q$ depends on the jet clustering algorithm.
The contribution from the jet function is proportional to the leading order cross section and 
has been included in the following hard matrix $\mathbf{H}$.

The hard matrix in the color basis  of Eq.~(\ref{eq:basis}) can be expressed as
\begin{eqnarray}
\mathbf{H}=  \left[
             \begin{array}{cc}
               H^{(0)}+H^{(1)} & H_{12}^{(1)} \\
               H_{12}^{(1)} & 0 \\
             \end{array}
           \right] \ ,
\end{eqnarray}
where $H^{(0)}$ denotes the leading order hard matrix element, which is given as
\begin{align}
H^{(0)}(ub\to dt)&=\dfrac{1}{C_A^2}\dfrac{g^4\hat{s}(\hat{s}-m_t^2)}{4(\hat{t}-m_W^2)^2}|V_{ud}|^2|V_{tb}|^2,
\end{align}
with $g$ being the  $SU(2)_L$ gauge coupling.
The CKM matrix element $V_{ij}$ needs to change in accordance with  the quark flavors of the  hard scattering process.
$m_W$ is the $W$-boson mass. The spin and color average factors have been included.
Another $t$-channel matrix element $H^{(0)}(\bar{u}b\to \bar{d}t)$ can be obtained by exchange $\hat{s}$ to $\hat{u}$.
The hard matrix elements $H^{(1)}$ and $H_{12}^{(1)}$ are from one-loop QCD correction. The numerical result shows that the  contribution from $H_{12}^{(1)}$ can be ignored and the analytical result can be obtained from Eq.~(A.7) of Ref.~\cite{Zhu:2010mr} by crossing symmetry (by exchanging $\hat{s}\leftrightarrow \hat{t}$ for $ub\to dt$ production process), thus we only show the diagonal element $H^{(1)}$ as below.
\begin{align}
H^{(1)}&=\dfrac{\alpha_s}{2\pi}H^{(0)}\left[-\ln^2(1-\lambda)-\dfrac{\ln(1-\lambda)}{\lambda}-2\ln(1-\lambda)
-2\ln(1-\lambda)\ln\dfrac{\hat{s}}{m_t^2}\right.\nn\\
&+\ln\dfrac{\mu_{\rm Res}^2}{\hat{s}}
\left(-2\ln(1-\lambda)-\ln\dfrac{\hat{s}}{m_t^2}-2\ln\dfrac{-\hat{s}}{\hat{t}}-\dfrac{11}{2}
\right)-\dfrac{1}{2}\ln^2\dfrac{\hat{s}}{m_t^2}-\dfrac{5}{2}\ln\dfrac{\hat{s}}{m_t^2} 
-\dfrac{3}{2}\ln\dfrac{P_{J\perp}^2R^2}{\mu_{\rm Res}^2}\nn\\
&\left.+2\mathrm{Li}_2(\lambda)+\dfrac{1}{2}\ln^2\dfrac{P_{J\perp}^2R^2}{\mu_{\rm Res}^2}
-\dfrac{3}{2}\ln^2\dfrac{\mu_{\rm Res}^2}{\hat{s}}-\ln^2\dfrac{-\hat{s}}{\hat{t}}-3\ln\dfrac{-\hat{s}}{\hat{t}}-\dfrac{5\pi^2}{6}-\dfrac{15}{2}
\right]
+\delta H^{(1)} ,
\end{align}
where $\lambda=\hat{t}/(\hat{t}-m_t^2)$, and
the $\delta H^{(1)}$ is not proportional to the leading order cross section,
\begin{equation}
\delta H^{(1)}=\dfrac{\alpha_s}{2\pi}\dfrac{1}{4C_A^2}\dfrac{g^4C_Fm_t^2}{(\hat{t}-m_W^2)^2}\dfrac{\hat{s}\hat{u}}{\hat{t}}\ln\dfrac{m_t^2}{m_t^2-\hat{t}}|V_{ud}|^2|V_{tb}|^2,
\end{equation}
where the spin and color average factors have also been included.

We should note that the non-global logarithms (NGLs) could also contribute to this process. The NGLs arise from some special kinematics of two soft gluon radiations, in which the first one is radiated outside of the jet which subsequently radiates a second gluon into the jet~\cite{Dasgupta:2001sh,Dasgupta:2002bw,Banfi:2003jj,Forshaw:2006fk}. Numerically, the NGLs are negligible in this process since it starts at $\mathcal{O}(\alpha_s^2)$. Therefore we will ignore their contributions in the following phenomenology discussion.

\section{Phenomenology}
\begin{figure}
\centering
\includegraphics[width=0.44\textwidth]{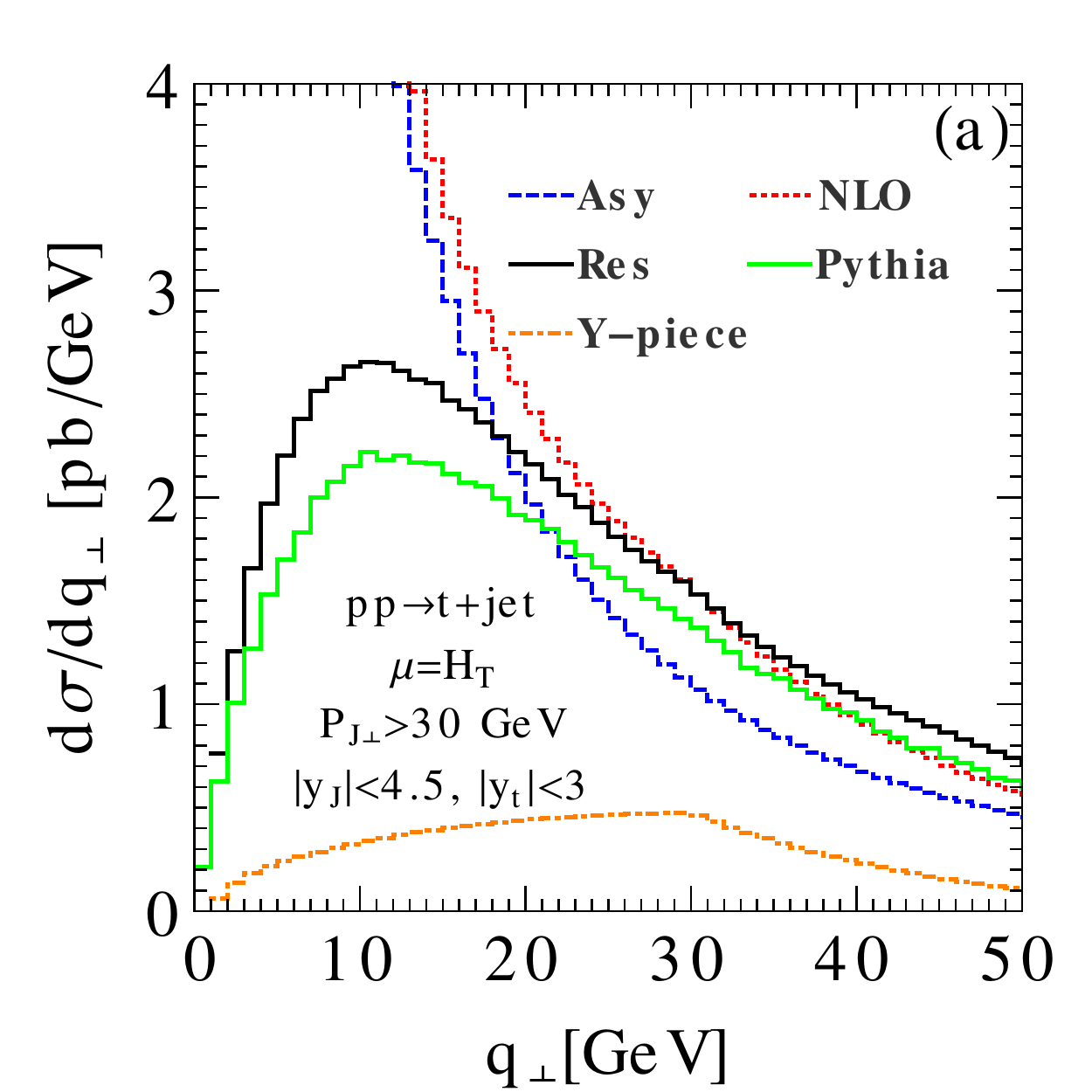}
\includegraphics[width=0.44\textwidth]{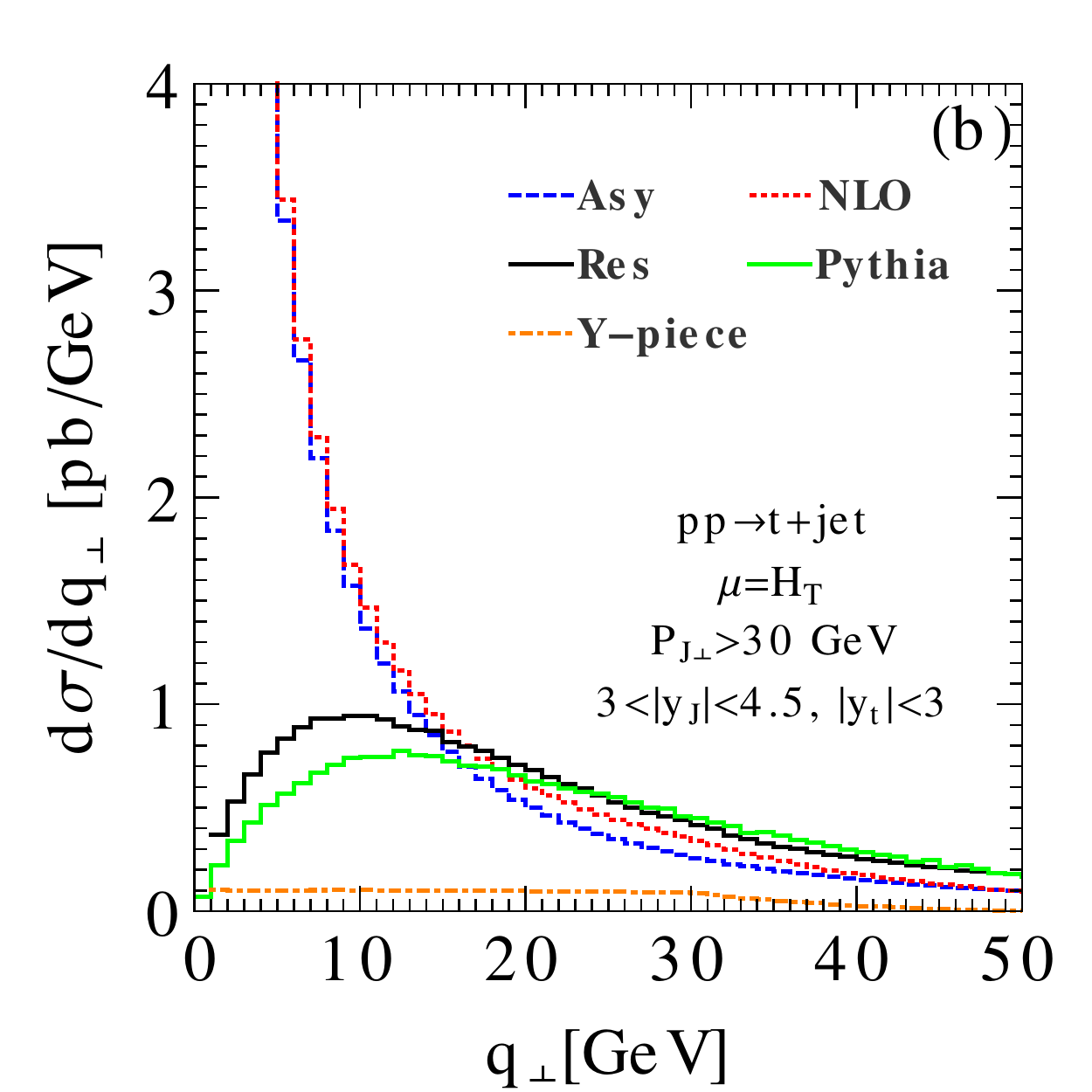}
\includegraphics[width=0.44\textwidth]{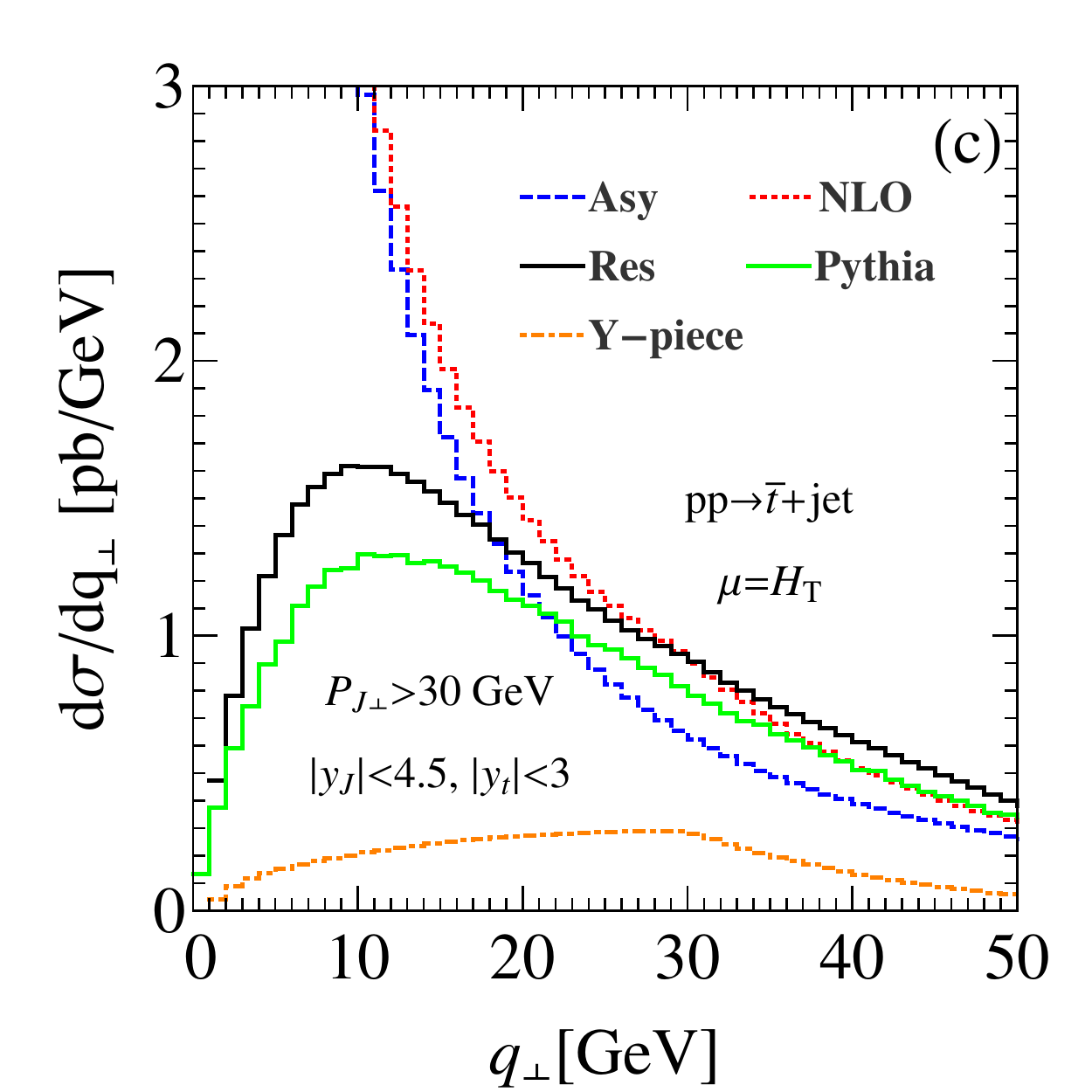}
\includegraphics[width=0.44\textwidth]{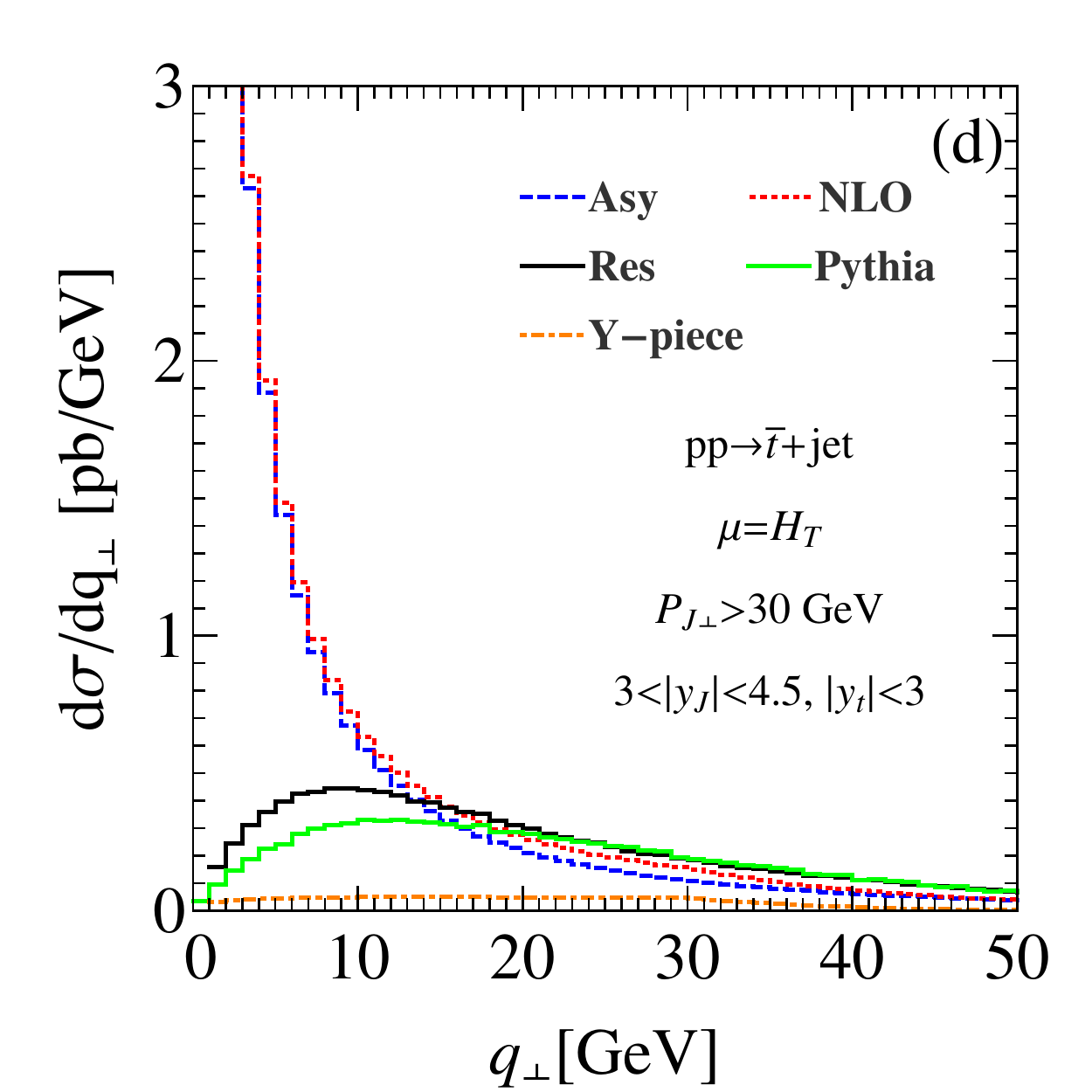}
\caption{ The $q_{\perp}$ distribution from the asymptotic result (blue dashed line), NLO calculation (red dotted line), resummation prediction (black solid line), parton shower result by PYTHIA 8 (green solid  line) and $Y$-term (orange dot-dashed  line) for the $t$-channel single top quark production (a, b) and the anti-top quark production (c, d)  at the $\sqrt{S}=13~{\rm TeV}$ LHC with $|y_t|<3$ and $|y_J|\leq 4.5$ (a, c), or $3.0\leq|y_J|\leq 4.5$ (b, d) . The resummation and renormalization scales are choose as $\mu=\mu_{\rm Res}=\mu_{\rm ren}=H_T$. }
\label{fig:tqt13}
\end{figure}

\begin{figure}
\centering
\includegraphics[width=0.44\textwidth]{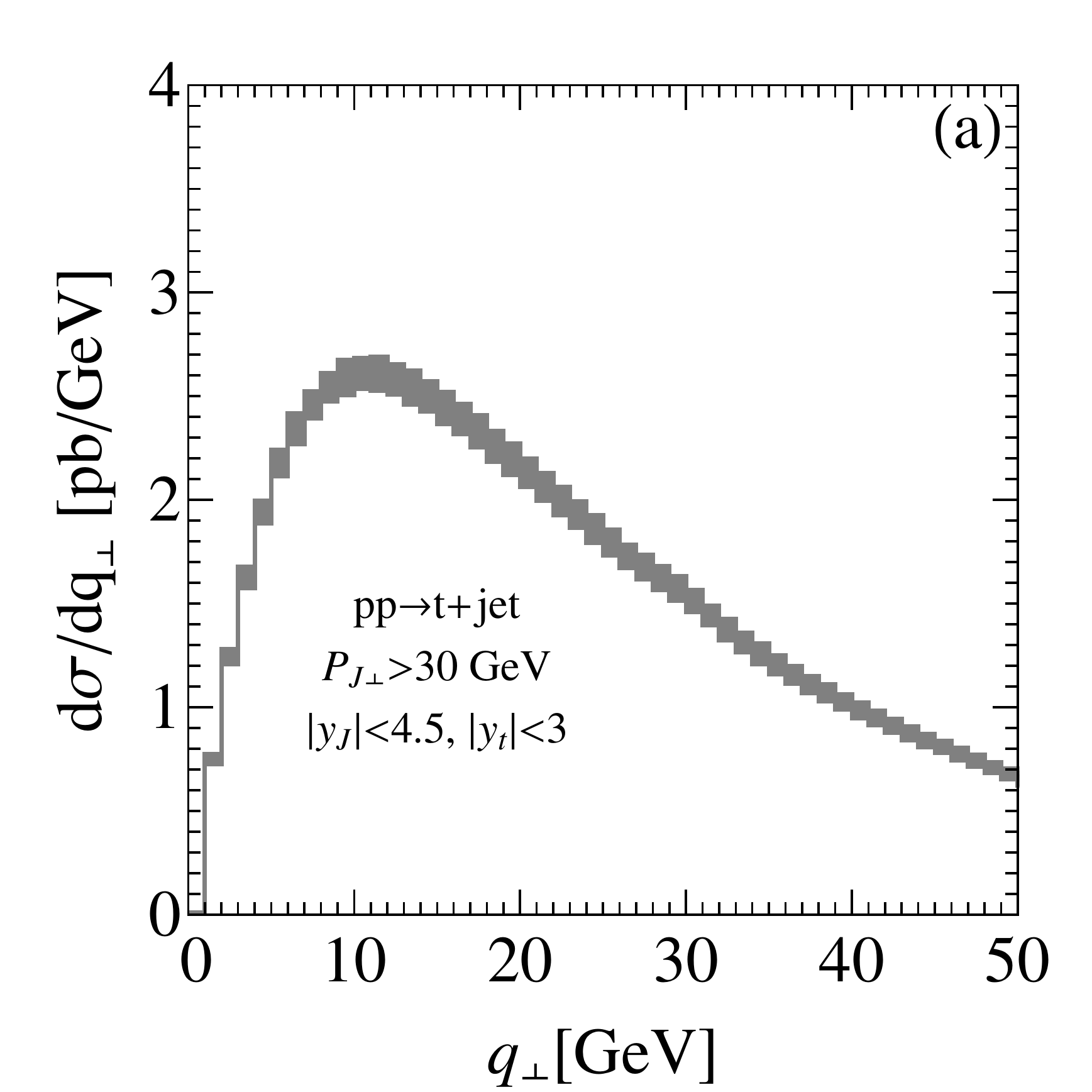}
\includegraphics[width=0.44\textwidth]{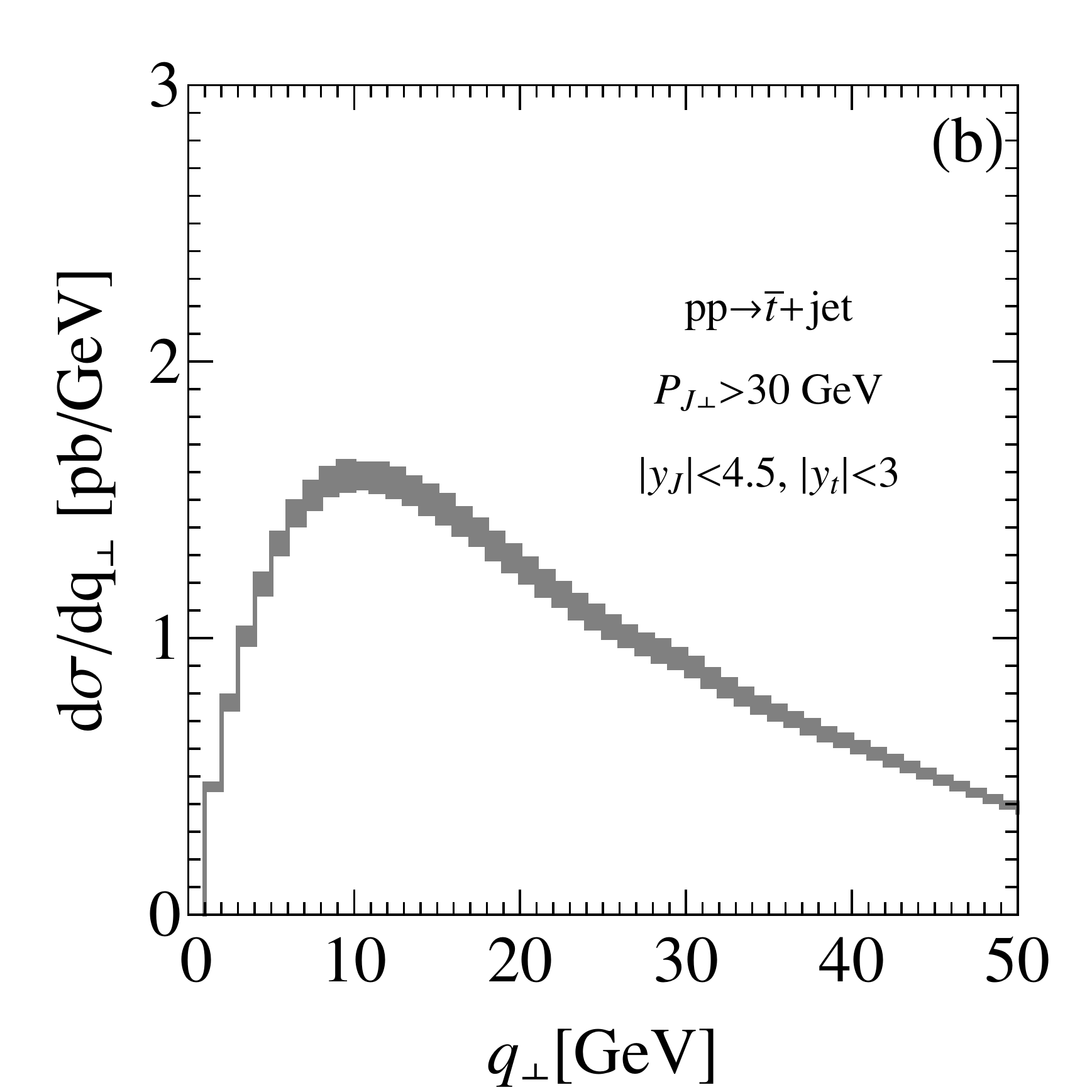}
\caption{The scale uncertainties for the $t$-channel single top quark production (a) and the anti-top quark production (b) at the $\sqrt{S}=13~{\rm TeV}$ LHC with $|y_t|<3$,  $|y_J|\leq 4.5$ and $P_{J\perp}>30~{\rm GeV}$. The resummation and renormalization scales are varied from $H_T/2$ to $2H_T$.}
\label{fig:scale}
\end{figure}

\begin{figure}
\centering
\includegraphics[width=0.44\textwidth]{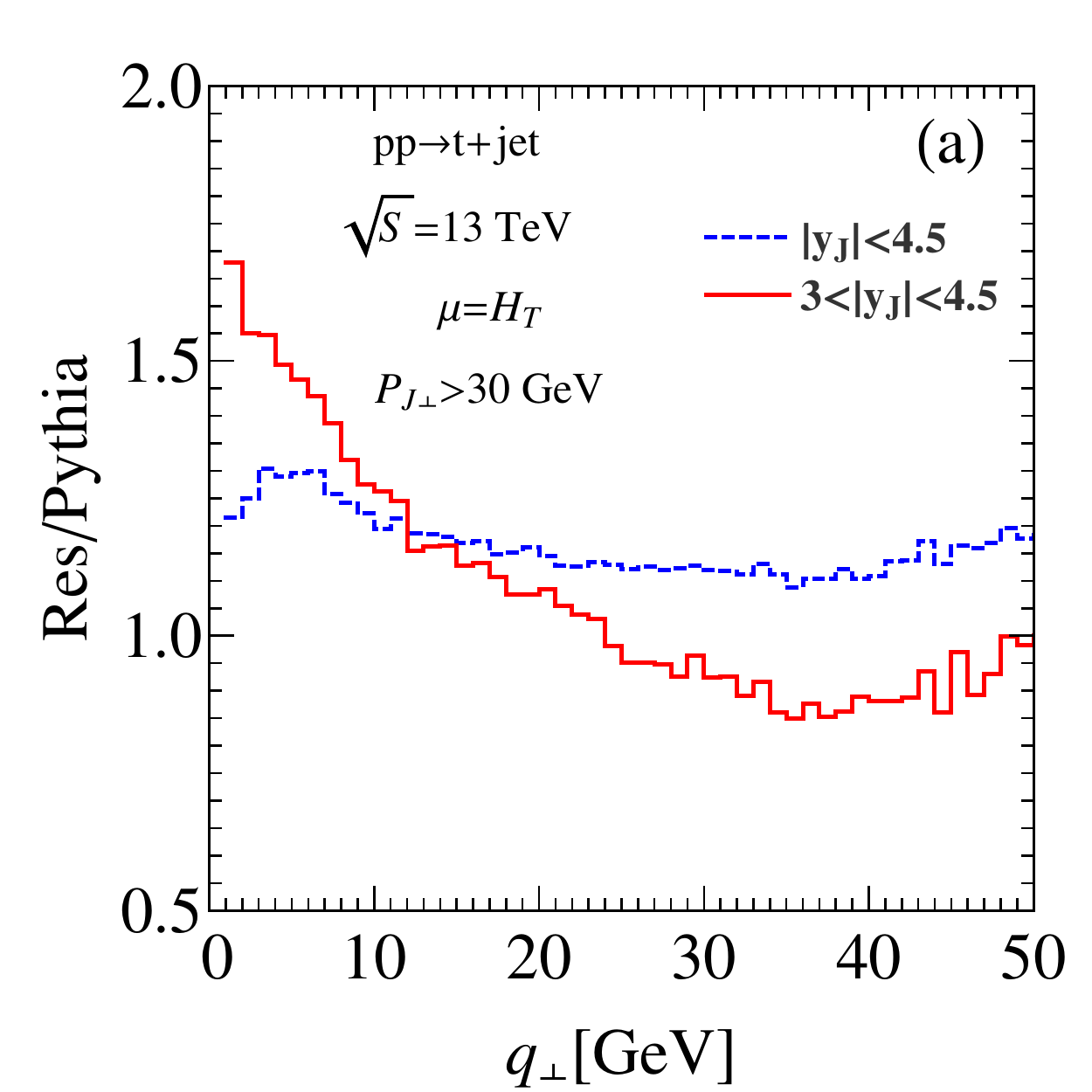}
\includegraphics[width=0.44\textwidth]{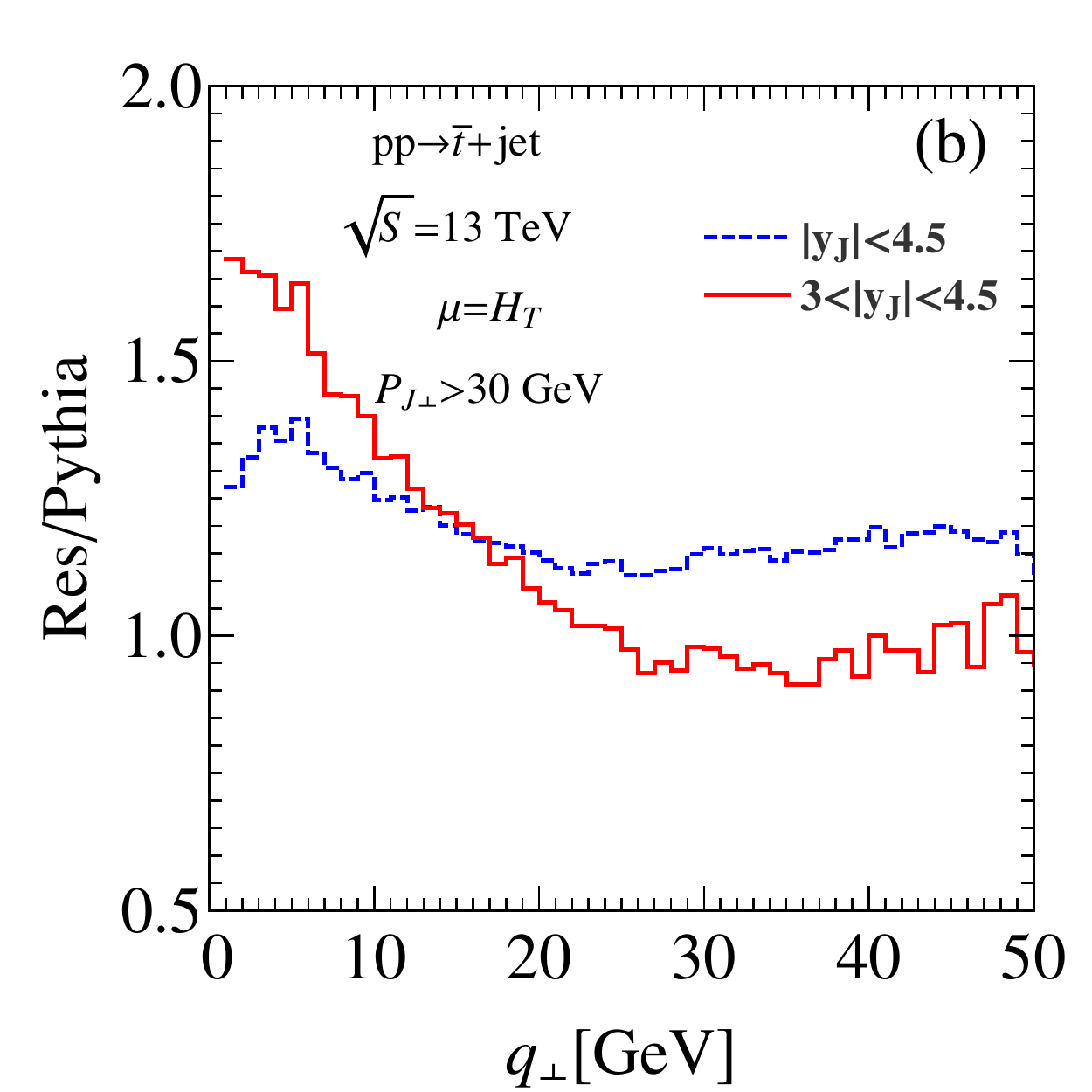}
\caption{The ratio of the resummation and PYTHIA prediction for the $t$-channel single top quark production (a) and anti-top quark production (b) at the $\sqrt{S}=13~{\rm TeV}$ LHC with $|y_t|<3$, $P_{J\perp}>30~{\rm GeV}$  and $|y_J|\leq 4.5$ (blue dashed line), or $3.0\leq|y_J|\leq 4.5$ (red solid line). The resummation and renormalization scales are choose as $\mu=\mu_{\rm Res}=\mu_{\rm ren}=H_T$. }
\label{fig:ratio}
\end{figure}
\begin{figure}
	\centering
	\includegraphics[width=0.44\textwidth]{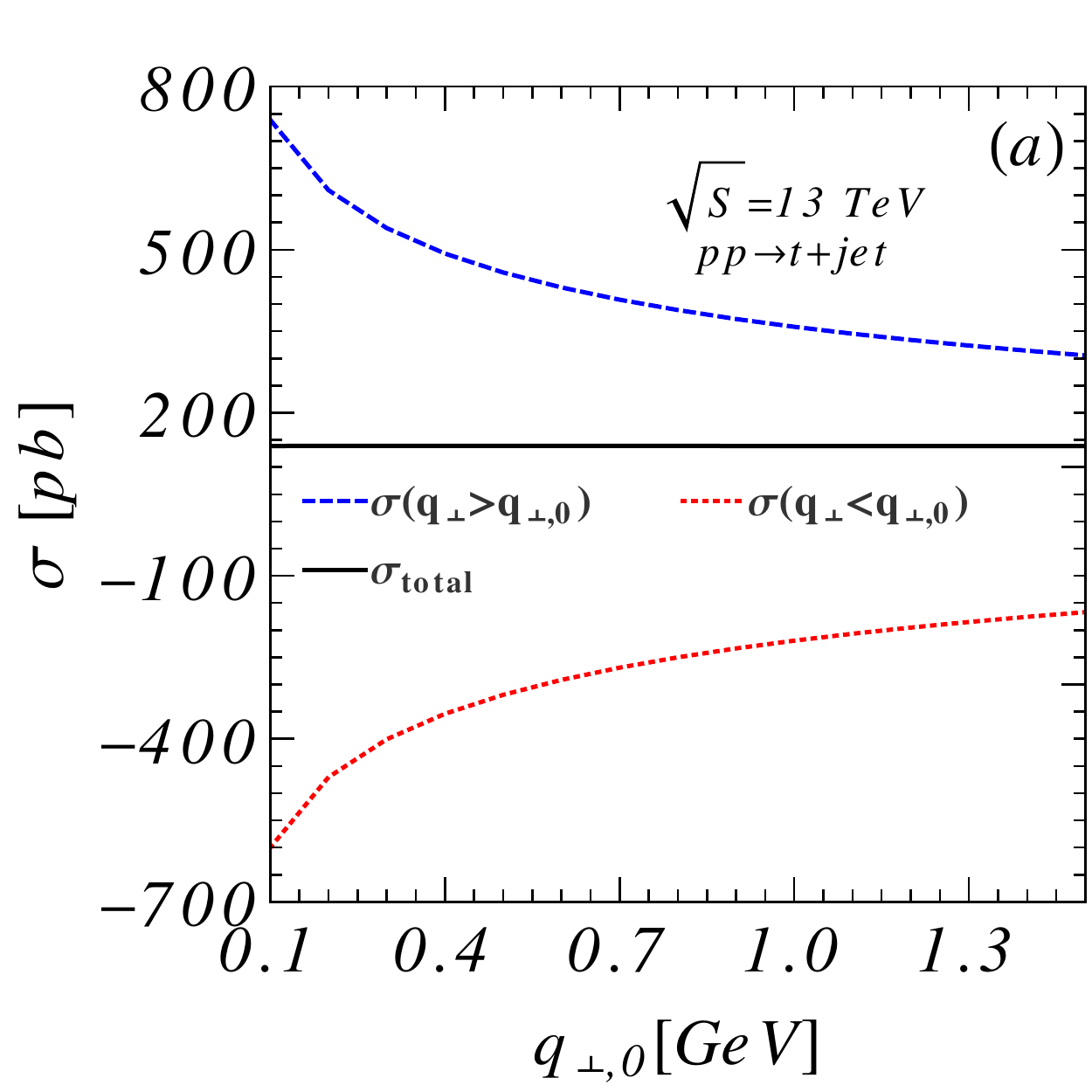}
	\includegraphics[width=0.44\textwidth]{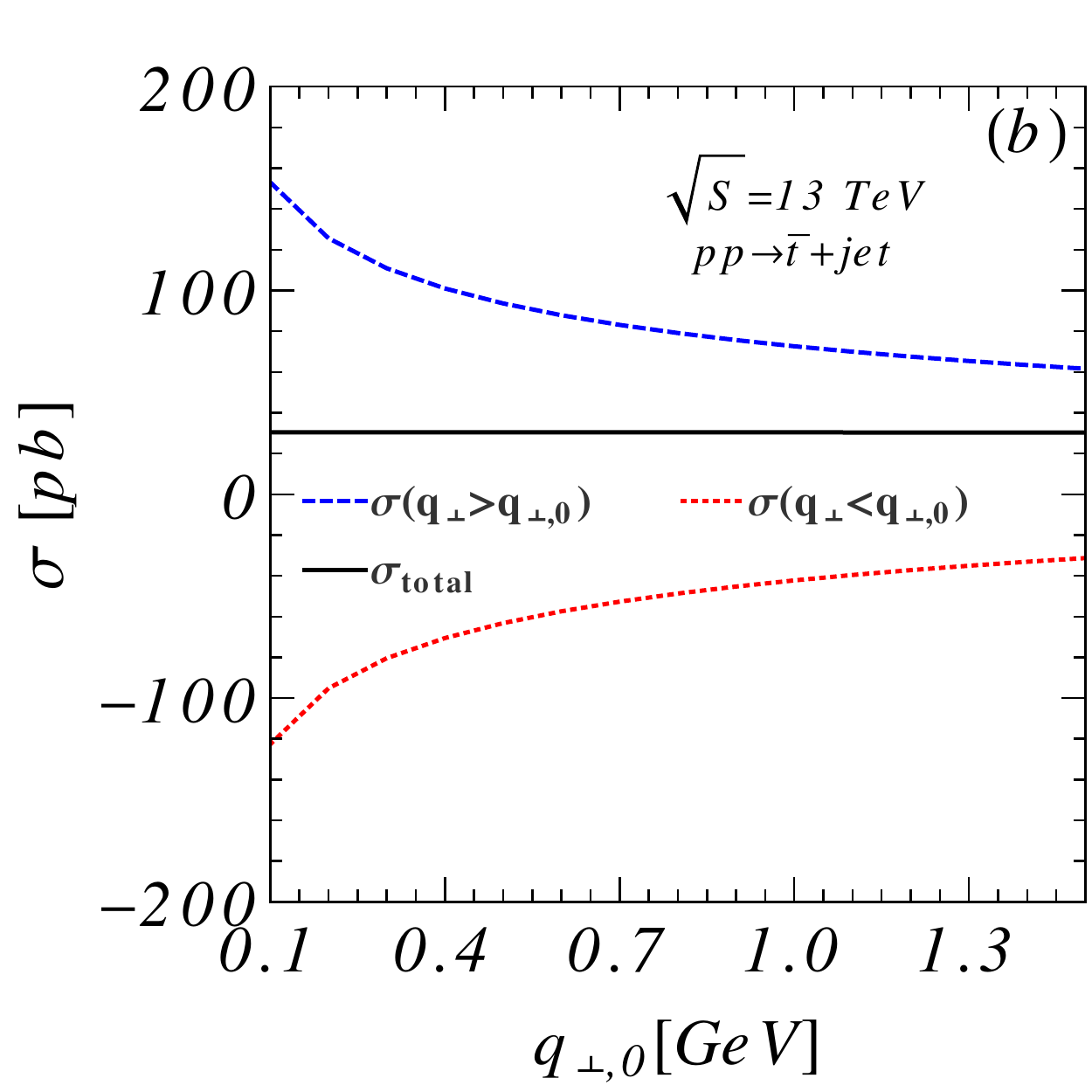}
	\includegraphics[width=0.44\textwidth]{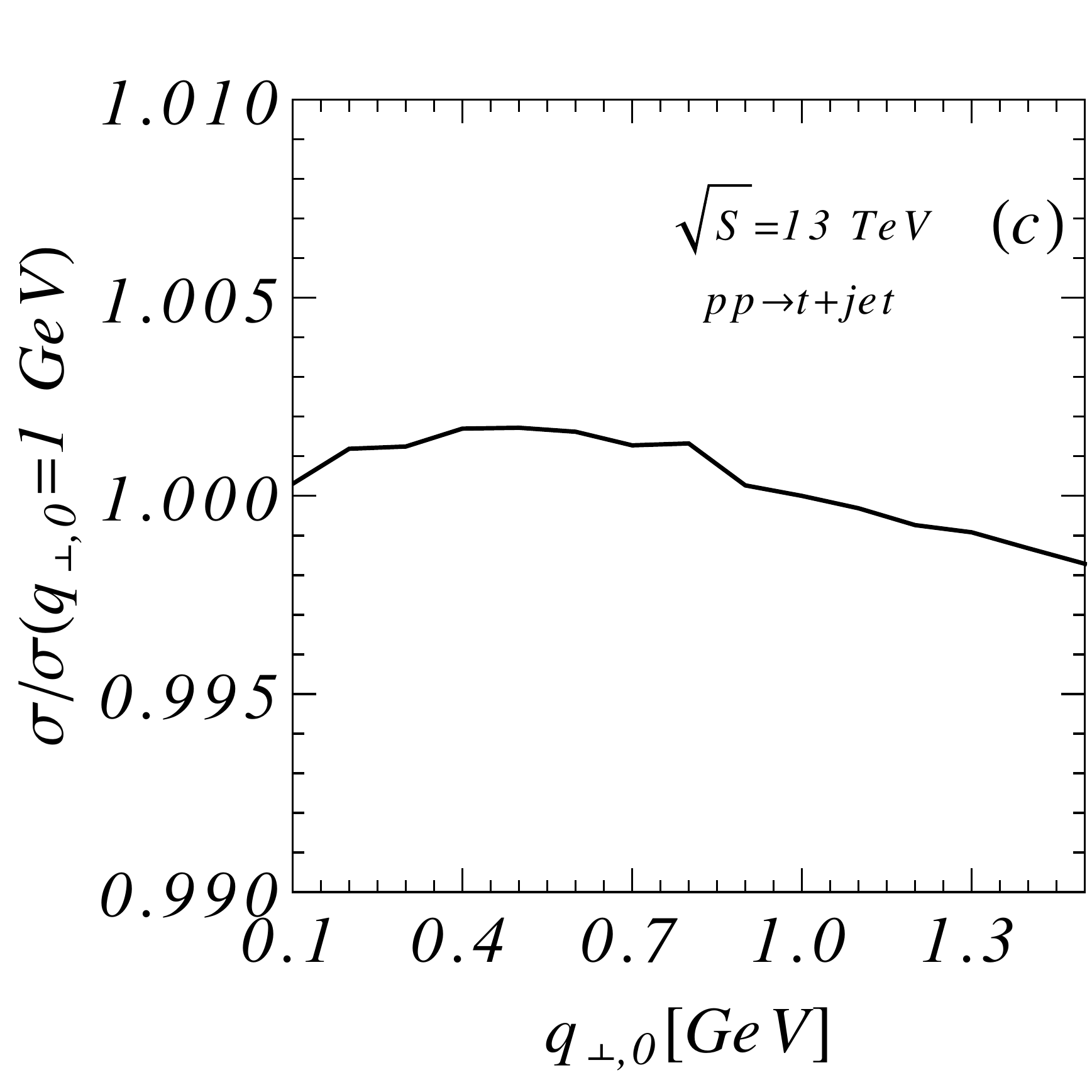}
	\includegraphics[width=0.44\textwidth]{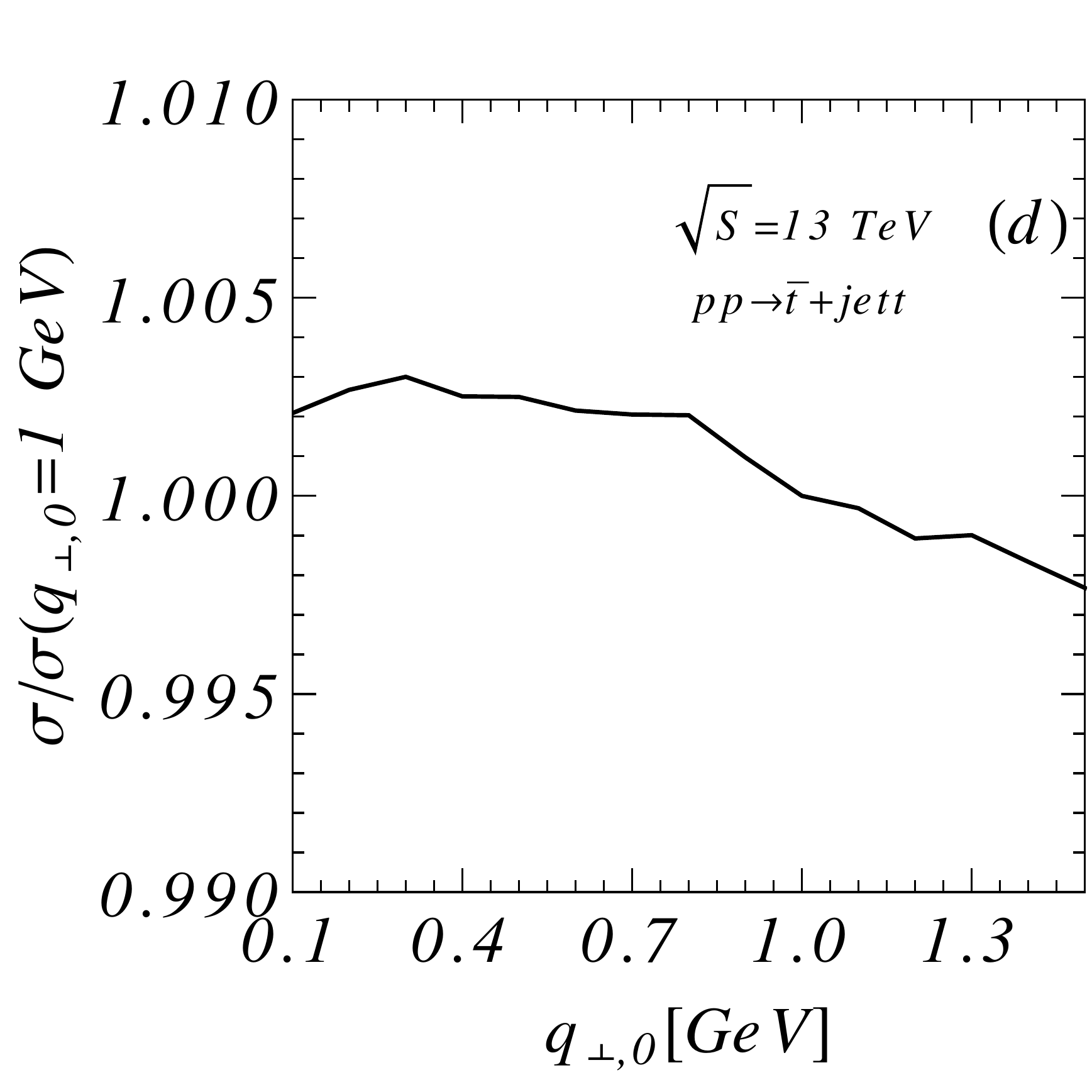}
	\caption{The upper blue dashed curve shows the contribution from the second term of the right-hand side of Eq. (\ref{eq:cstot}), the lower red dotted curve from the first term, and the black solid line shows the total cross section, at the NLO (a, b).  The ratio plots $\sigma/\sigma(q_{\perp,0}=1~{\rm GeV})$ are shown in (c, d).  }
	\label{fig:cs}
\end{figure}

Below, we present the numerical result of resummation calculation for the
$t$-channel single top (anti-)  quark
production at the $\sqrt{S}=13~{\rm TeV}$ LHC with CT14NNLO PDF~\cite{Dulat:2015mca}.
Figure~\ref{fig:tqt13} shows the $q_{\perp}$ distribution from the asymptotic piece (blue dashed line), NLO calculation (red dotted  line), resummation prediction (black solid line) and $Y$-term (orange dot-dashed line) for the top quark (a, b) and the anti-top quark (c, d)  production. Here, the asymptotic piece is the fixed-order expansion of Eq.~(\ref{resumy}) up to the $\alpha_s$ order, and
is expected to agree with the NLO prediction as $q_{\perp} \to 0$.
In the same figure, we also compare to the prediction from the
parton shower event generator PYTHIA 8~\cite{Sjostrand:2007gs} (green solid  line),
which was calculated at the leading order, with CT14LO PDF.
For the fixed-order calculation, both the renormalization and factorization scales are fixed at
$H_{T}\equiv\sqrt{m_t^2+P_{J\perp}^2}+P_{J\perp}$.
Similarly, in the resummation calculation, the canonical choice of
the resummation ($\mu_{\rm Res}$) and renormalization ($\mu_{\rm ren}$) scales
is taken to be $H_T$ in this study.
The uncertainties of the resummation predictions are estimated by varying the scale $\mu_{\rm Res}=\mu_{\rm ren}$ by a factor two around the central value $H_T$ , which is shown in  Fig.~\ref{fig:scale}.
The jet cone size is choose as $R=0.4$,
using the anti-$k_T$ algorithm, and the Wolfenstein CKM matrix elements parameterization is used in our numerical calculation~\cite{Olive:2016xmw}.
We shall compare predictions for two different sets of kinematic cuts, with
$|y_t|\leq 3$ and $P_{J\perp}>30~{\rm GeV}$, and
 $|y_J|\leq 4.5$ in (a, c), and
$3\leq |y_J|\leq 4.5$ in (b, d) of Fig.~\ref{fig:tqt13},
respectively.
Some results of the comparison are in order.
Clearly, the asymptotic piece and the fixed-order calculation results agree very well in the small
$q_\perp$ (less than $1~{\rm GeV}$) region.
As a further check of our resummation calculation,
we integrate out the $q_{\perp}$ distribution to compare the total cross section with
that predicted by the NLO program MCFM~\cite{Campbell:2015qma}.
In the resummation framework, the NLO total cross section can be divided into two parts, the small $q_{\perp}$ region, which can be obtained by integrating the distribution of the asymptotic part and the
one-loop virtual diagram contribution,
and the large $q_{\perp}$ part, which is infrared safe and can be numerically calculated directly.
Thus, the total cross section is given by
\bea
\sigma_{NLO}=\int_0^{q_{\perp,0}^2}dq_{\perp}^2\dfrac{d\sigma_{NLO}^{virtual+real}}{dq_{\perp}^2}+\int_{q_{\perp,0}^2}^{\infty}dq_{\perp}^2\dfrac{d\sigma_{NLO}^{real}}{dq_{\perp}^2}.\nn\\
\label{eq:cstot}
\eea
It is obvious that the  two  contributions on the right-hand side side of the above equation depend on the cut-off parameter $q_{\perp,0}$ individually, but their sum is indepdendt of it.  We show the total NLO cross section of $pp\to t (\bar{t})+jets$ as a function of $q_{\perp,0}$,  in Fig.~\ref{fig:cs}. It is clear that the total cross sections of top quark and anti-top quark do not depend on  $q_{\perp,0}$, which  varies from 0.1 GeV to 1.5 GeV. 
We also  checked that the total cross sections
from our resummation and MCFM calculations are in perfect agreement.

\begin{figure}
\centering
\includegraphics[width=0.44\textwidth]{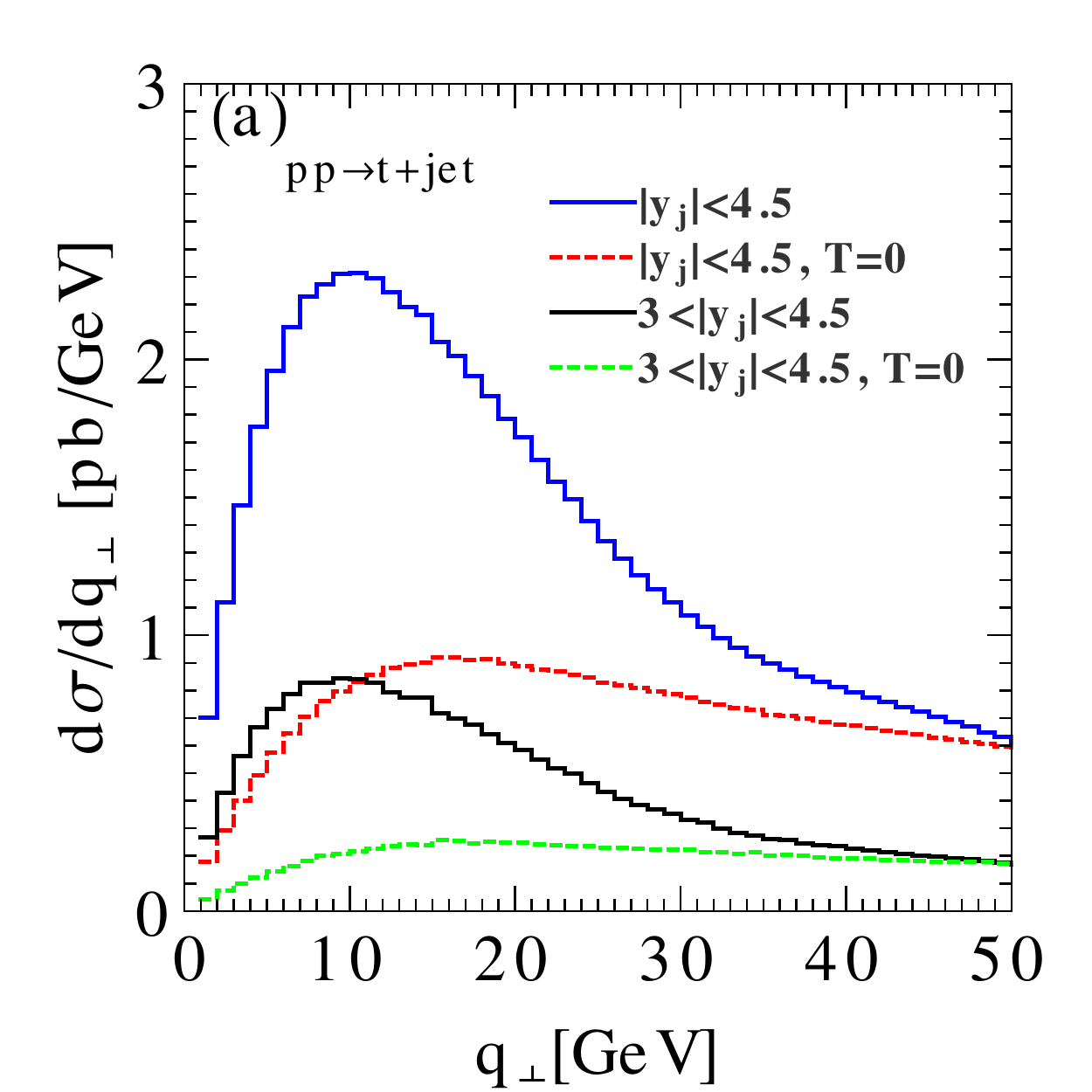}
\includegraphics[width=0.44\textwidth]{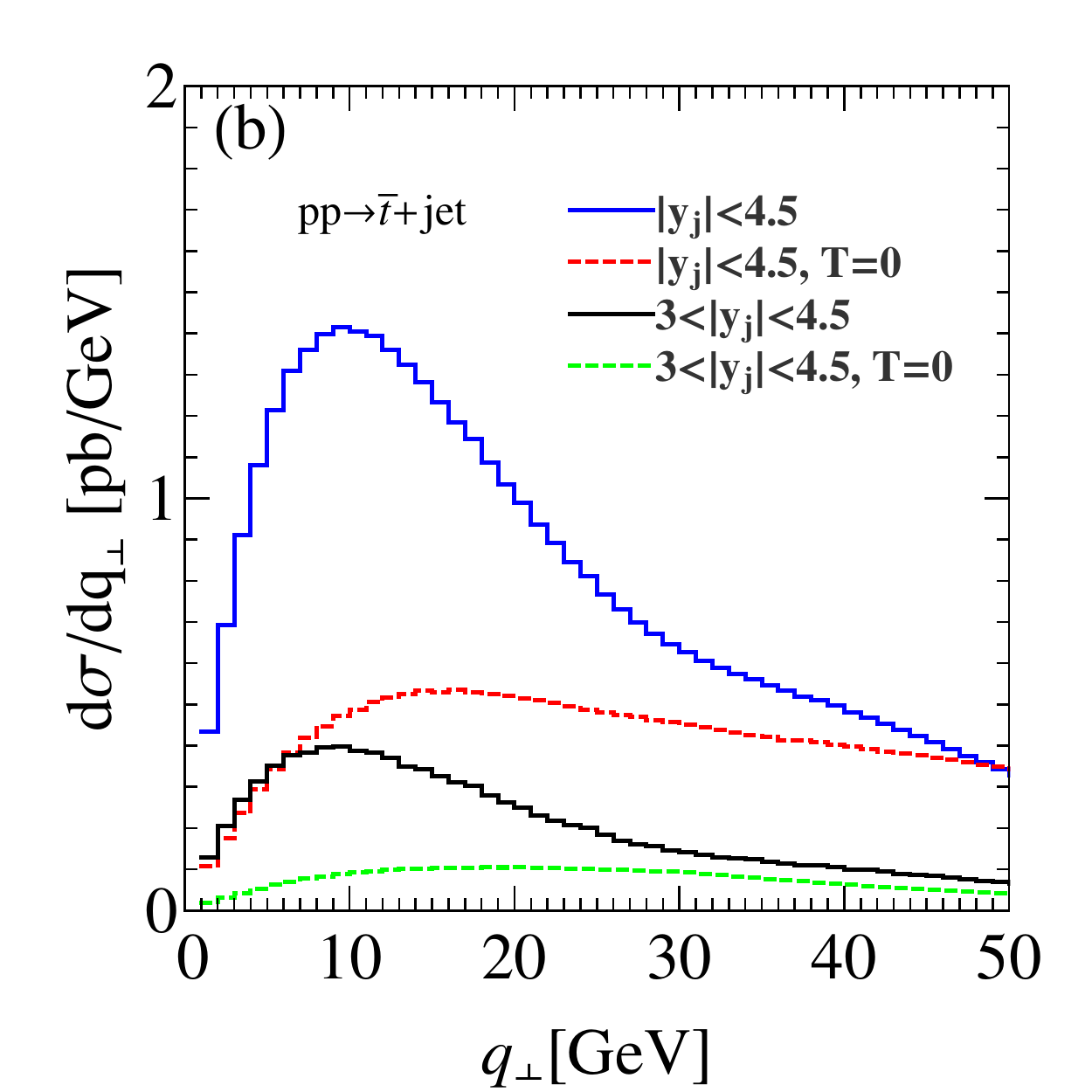}
\caption{(a)The  $W$-piece prediction  for the $t$-channel single top quark production (a) and anti-top quark production (b) at the $\sqrt{S}=13~{\rm TeV}$ LHC with  $|y_t|<3$ and $P_{J\perp}>30~{\rm GeV}$.
The blue solid and red dashed line represents the prediction with and without including the factor  $T$ in Eqs.~(\ref{sud12})-(\ref{sud15}) with $|y_J|\leq 4.5$, respectively, black solid and green dashed lines are for the $3\leq|y_J|\leq 4.5$. The resummation and renormalization scales are choose as $\mu=\mu_{\rm Res}=\mu_{\rm ren}=H_T$.}
\label{fig:tqt13nt}
\end{figure}

\begin{figure}
\centering
\includegraphics[width=0.44\textwidth]{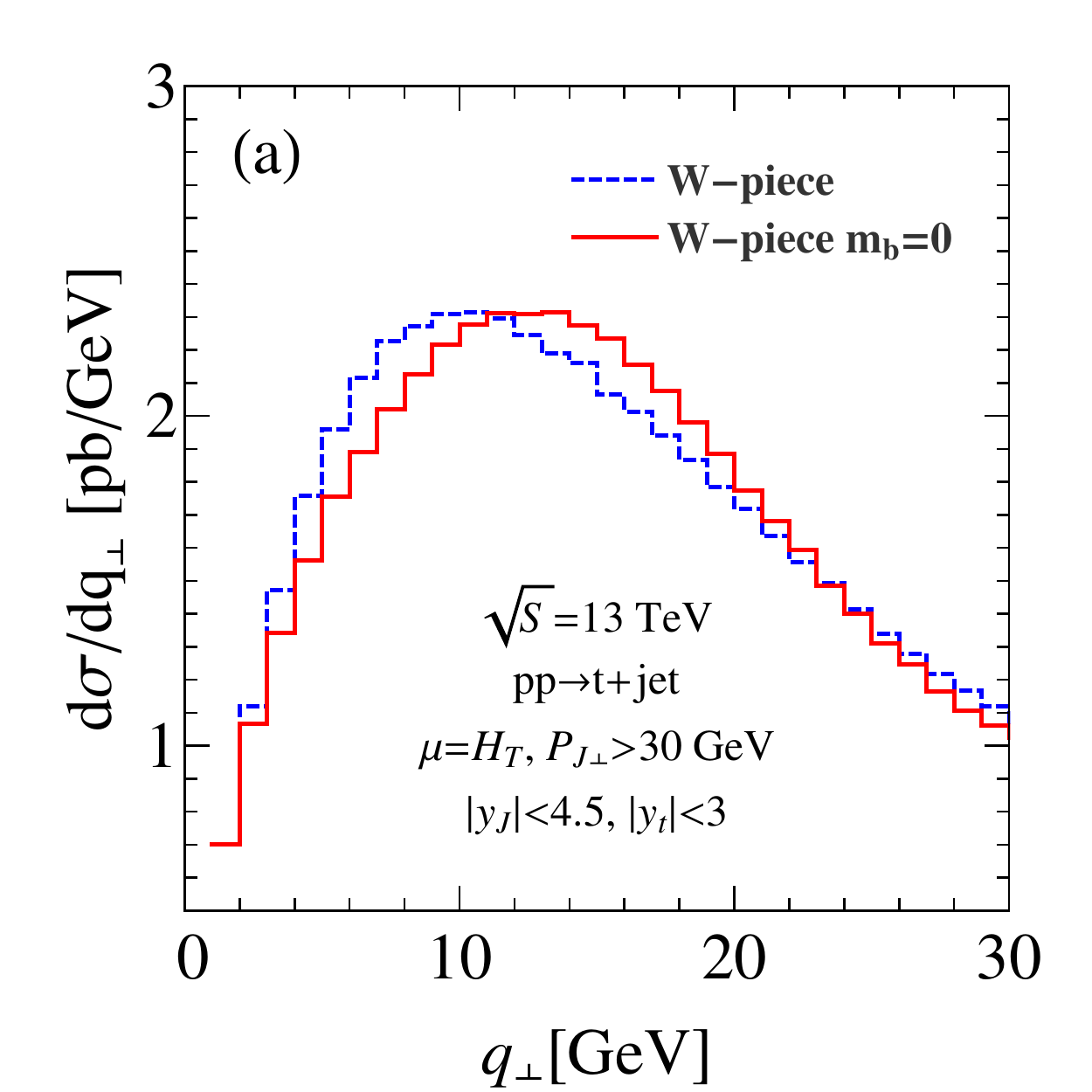}
\includegraphics[width=0.44\textwidth]{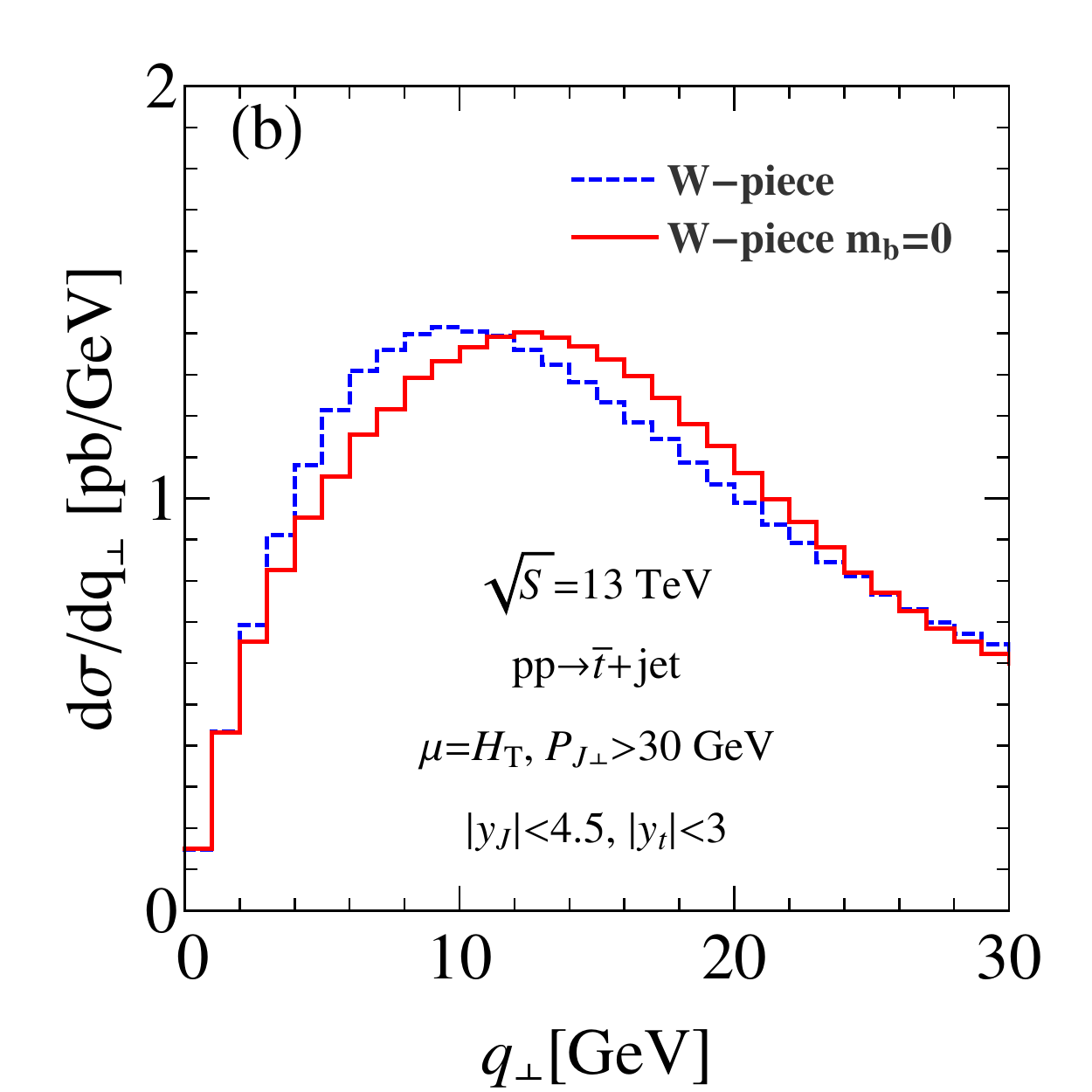}
\caption{ The $W$-piece prediction for the single top quark production (a) and anti-top quark production with non-zero $m_b$ (blue dashed  line ) and $m_b=0$ (red solid line) at the $\sqrt{S}=13~{\rm TeV}$ LHC with $|y_J|\leq 4.5$, $|y_t|<3$ and $P_{J\perp}>30~{\rm GeV}$. The resummation and renormalization scales are choose as $\mu=\mu_{\rm Res}=\mu_{\rm ren}=H_T$. }
\label{fig:nb}
\end{figure}

As shown in Fig.~\ref{fig:tqt13}, the NLO prediction is not reliable when the $q_{\perp}$ is small. The resummation calculation predicts a well behavior $q_{\perp}$ distribution in the small $q_{\perp}$ region since the large logarithms have been properly resummed.
In Fig~\ref{fig:ratio}, we compare the predictions from our resummation calculation to PYTHIA by taking the ratio of
their $q_{\perp}$ differential distributions shown in Fig.~\ref{fig:tqt13}.
With the jet rapidity $|y_J|\leq 4.5$ (blue solid line), its ratio is not sensitive to $q_{\perp}$,
for either single top (a) or anti-top quark (b)  production.
Hence, they predict almost the same shape in $q_{\perp}$ distribution, while
they predict different fiducial total cross sections because PYTHIA prediction includes only
leading order matrix element and is calculated with CT14LO PDFs.
However, if we require the final state jet to be in the forward rapidity region,
with $3\leq |y_J|\leq 4.5$ (red solid line), which is the so-called signal region of single top events~\cite{Aaboud:2017pdi},
we find that PYTHIA prediction disagrees with our resummation calculation.
Our resummation calculation predicts a smaller $q_{\perp}$ value when the final state
jet is required to fall into the forward region, i.e., the signal region.
We have checked that the PYTHIA result is not sensitive to the effects from
beam remnants. Furthermore,
the $Y$-term contribution, from NLO, is negligible in this region, cf.
Fig.~\ref{fig:tqt13}(b) and (d) (orange dot-dashed line).
Hence, we conclude that their difference most likely comes from the
treatment of multiple soft gluon radiation.

\begin{figure}
\centering
\includegraphics[width=0.44\textwidth]{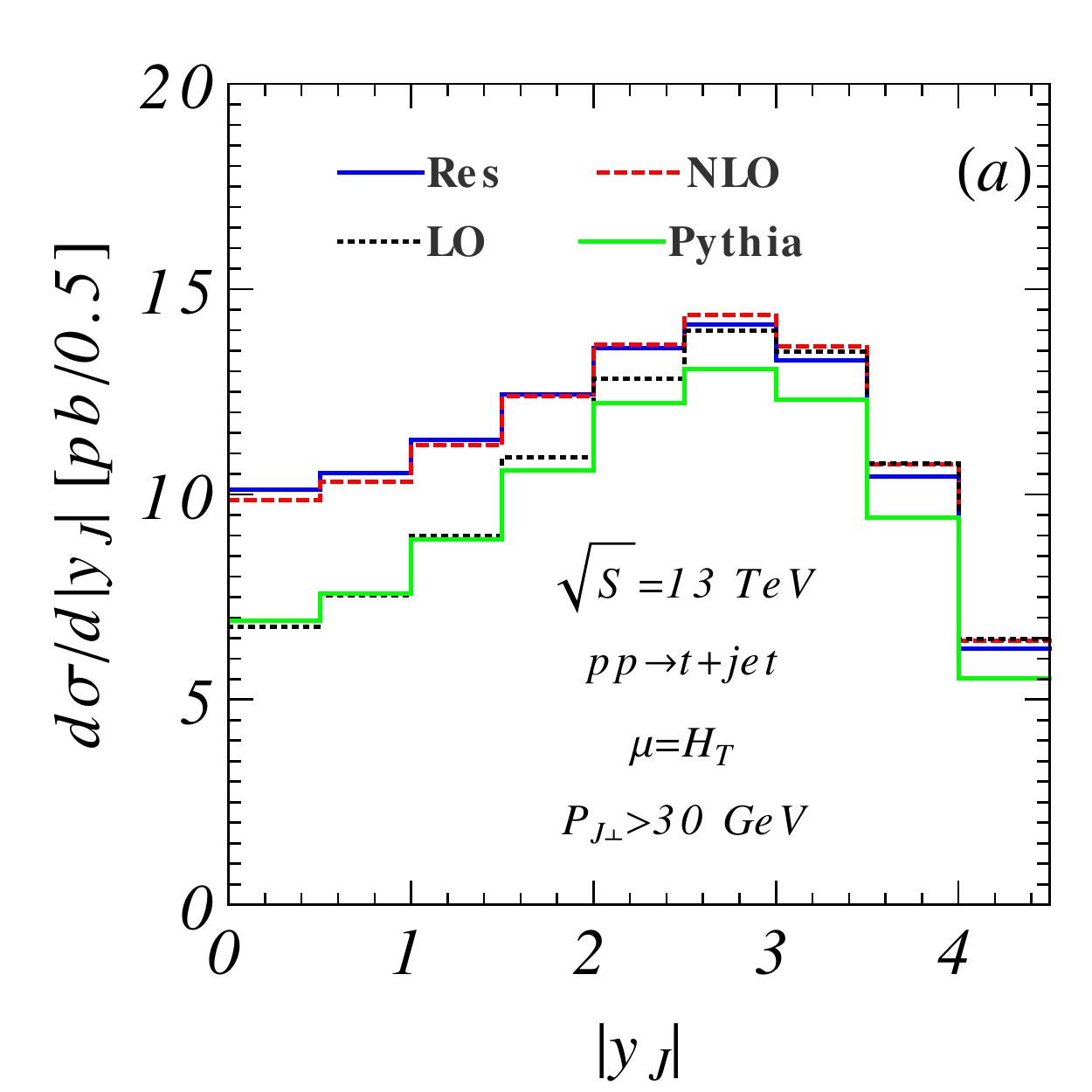}
\includegraphics[width=0.44\textwidth]{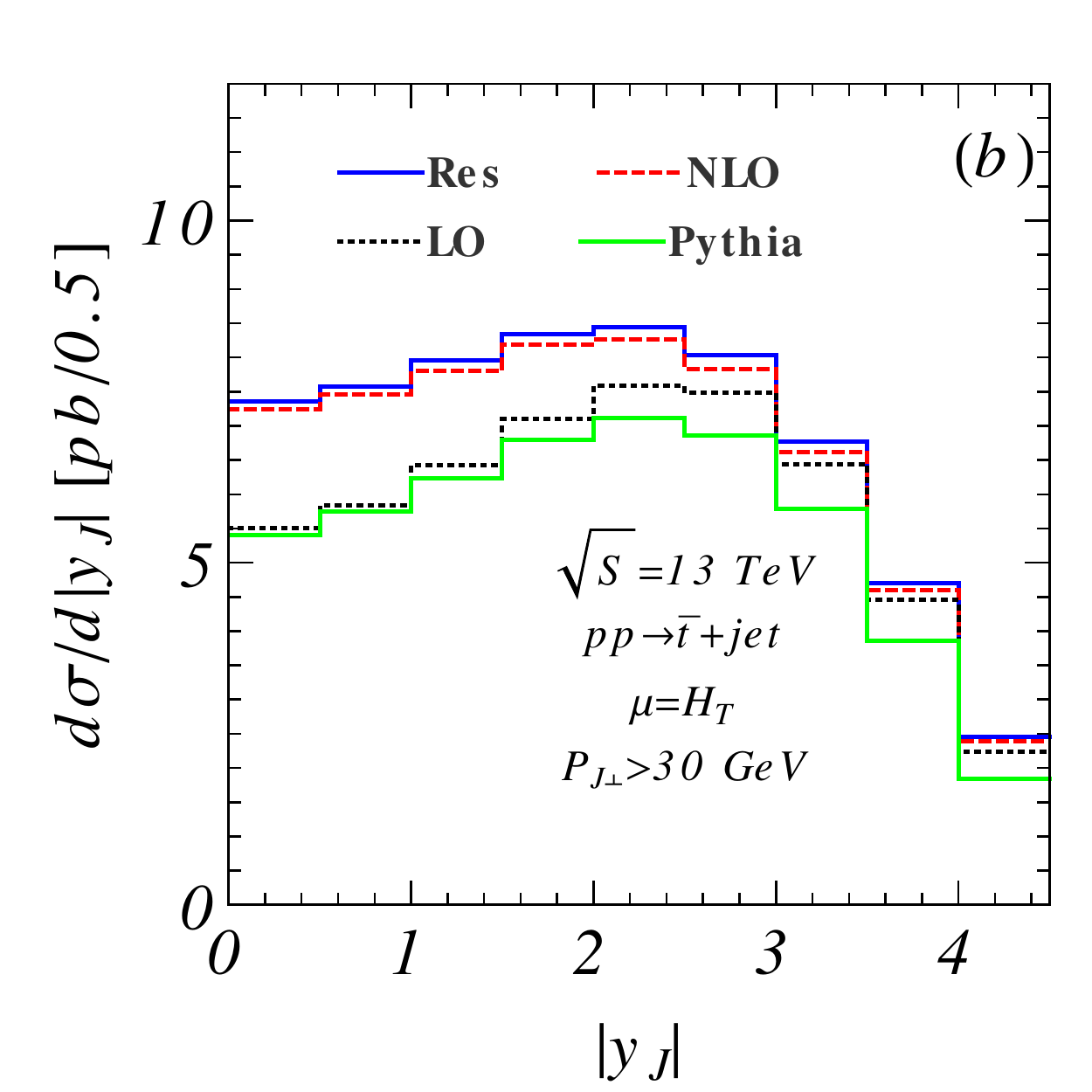}
\caption{(a) The jet rapidity distribution for the $t$-channel single top quark production (a) and anti-top quark production (b) from resummation prediction (blue solid line), NLO results (red dashed line), LO calculation with CT14LO PDF (black dashed line) and PYTHIA prediction (green solid line) at the $\sqrt{S}=13~{\rm TeV}$ LHC with $|y_t|<3$ and $P_{J\perp}>30~{\rm GeV}$.The renormalization and factorization scales are choose as $\mu=\mu_{\rm ren}=\mu_{\rm F}=H_T$ in the fixed-order calculation.
}
\label{fig:yj}
\end{figure}

\begin{figure}
\centering
\includegraphics[width=0.44\textwidth]{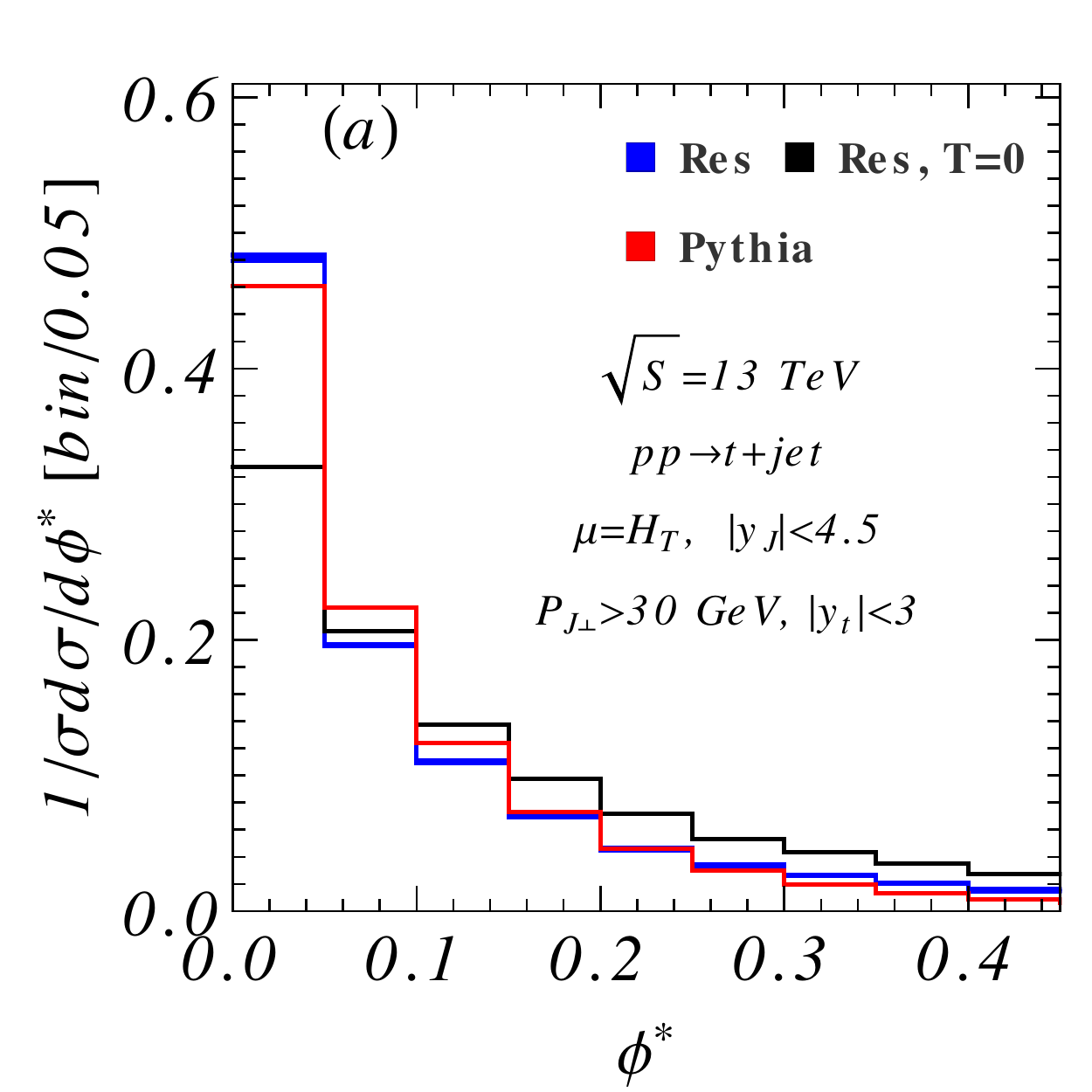}
\includegraphics[width=0.44\textwidth]{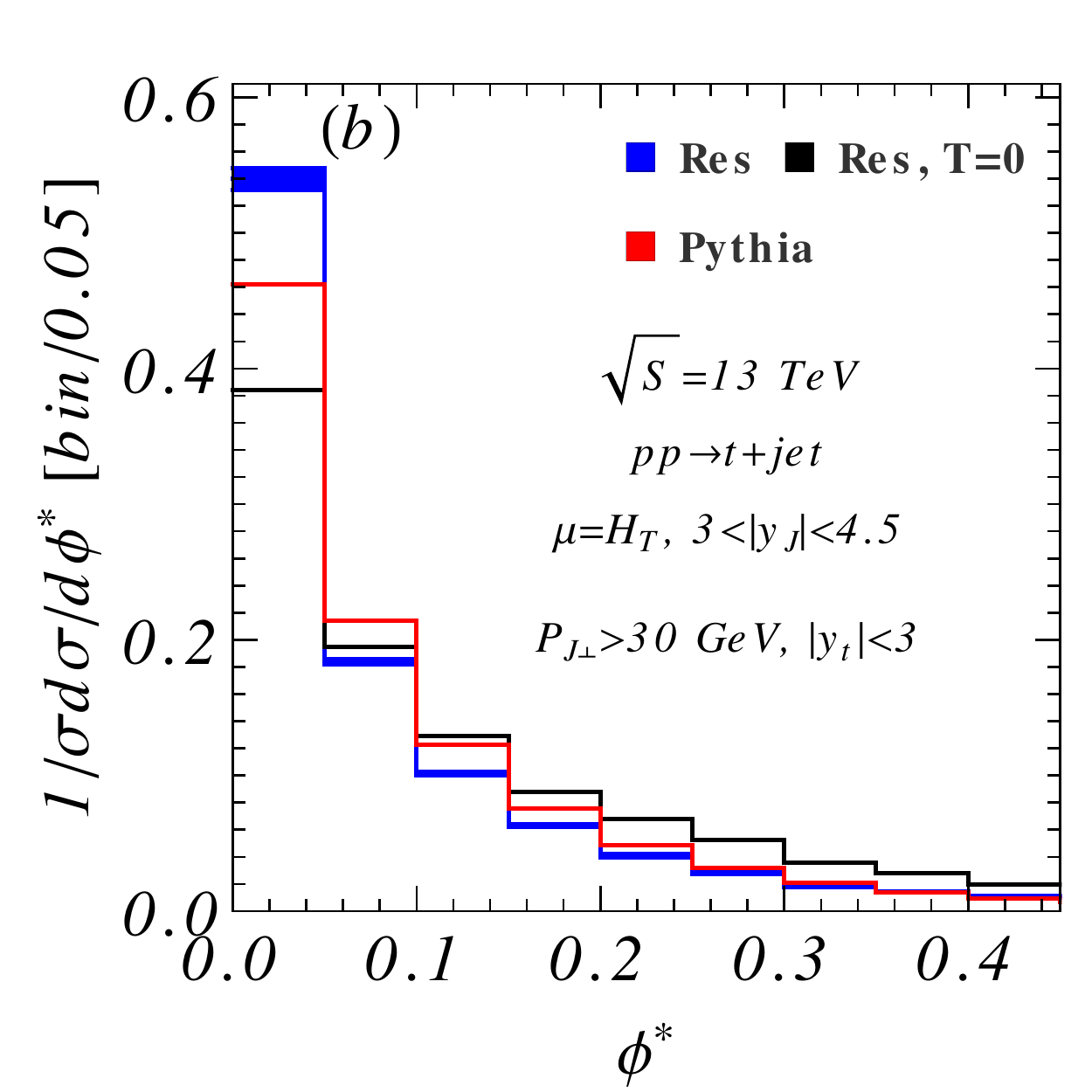}
\includegraphics[width=0.44\textwidth]{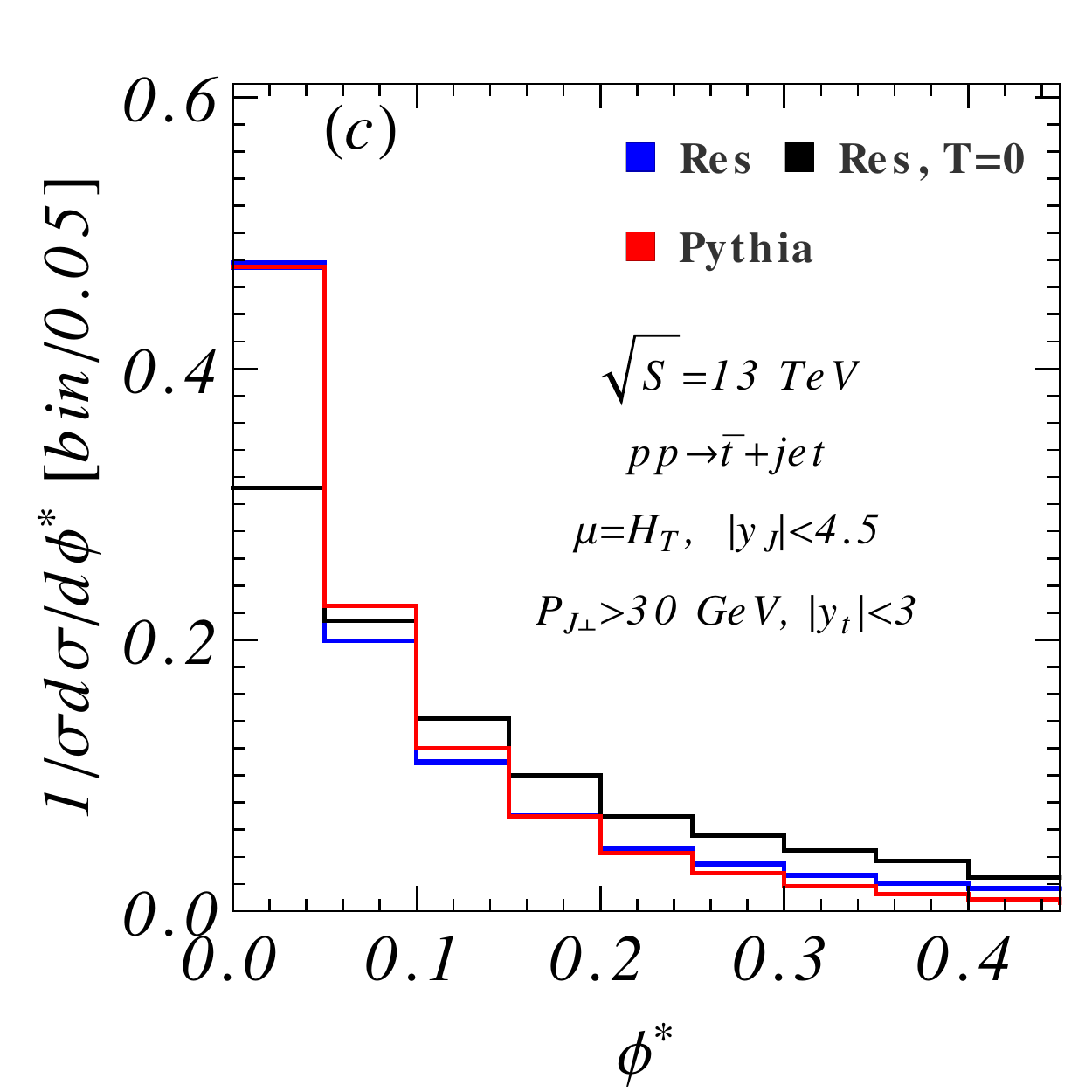}
\includegraphics[width=0.44\textwidth]{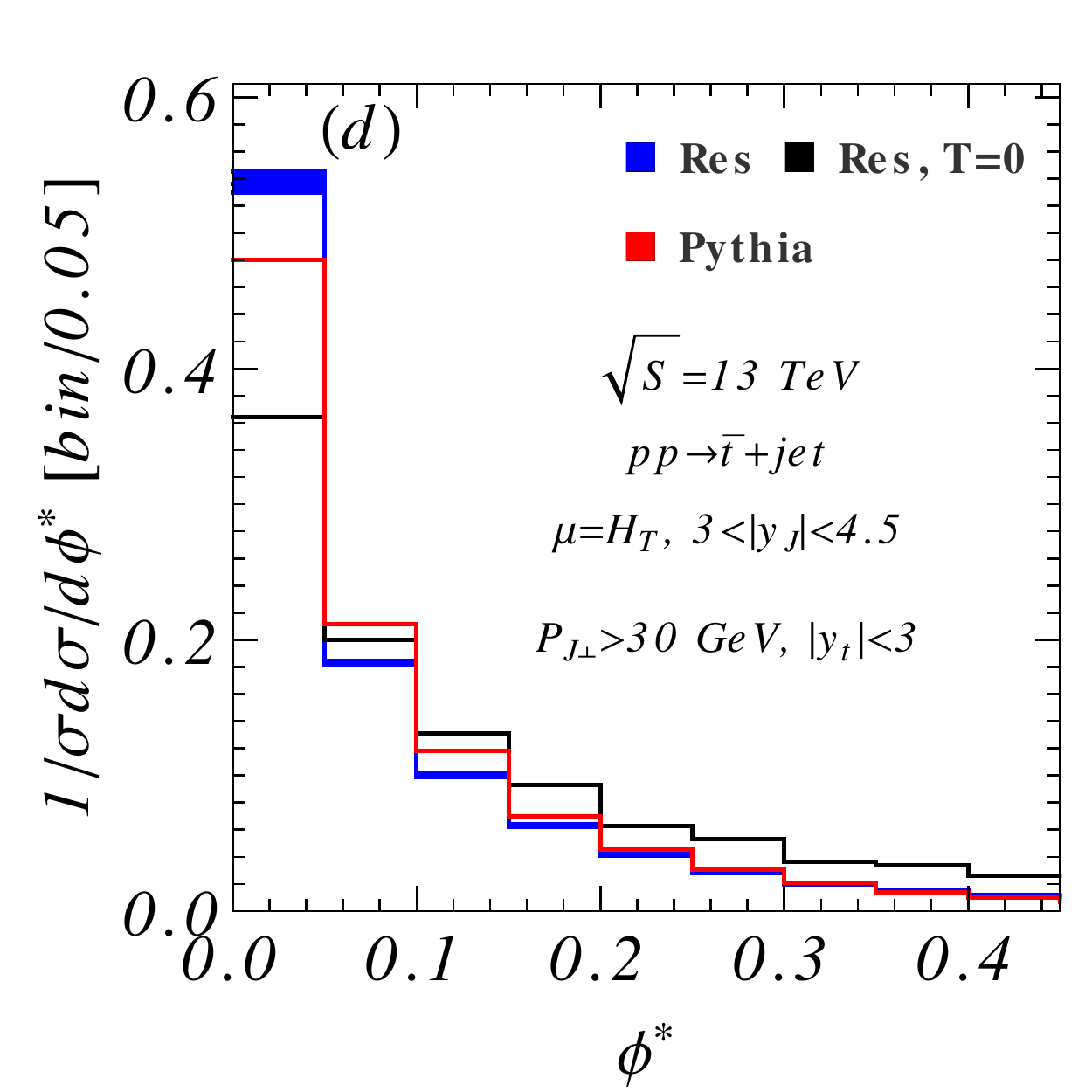}
\caption{The normalized distribution of $\phi^*$ for top quark production at the $\sqrt{S}=13~{\rm TeV}$ LHC with $|y_t|<3$ and $P_{J\perp}>30~{\rm GeV}$. The resummation and renormalization scales are choose as $\mu=\mu_{\rm Res}=\mu_{\rm ren}=H_T$. The blue and black line represents the resummation prediction with and without including the factor  $T$ in Eqs.~(\ref{sud12})-(\ref{sud15}), respectively. The  red lines describe the results from PYTHIA prediction. The blue shaded region represents the scale uncertainties which are varied from $H_T/2$ to $2H_T$.}
\label{fig:phi}
\end{figure}

As shown in Eqs.~(\ref{sud12})-(\ref{sud15}),
the effect of multiple gluon radiation, originated from
soft gluons connecting the initial and final state gauge links,
becomes more important when the final state jet is
required to be in the forward region where the
kinematic factor $T\sim \ln\dfrac{-\hat{t}}{\hat{s}}$
becomes large as $|\hat{t}| \to 0$. Consequently,
the $q_\perp$ distribution peaks at a smaller value  as  compared to the
case in which the final state jet does not go into the forward region.
To illustrate this, we compare in Fig.~\ref{fig:tqt13nt}
the predictions of the $W$-term in our resummation calculation, cf. the first term of Eq.~(\ref{resumy}),
with or
without including the factor $T$ in Eqs.~(\ref{sud12})-(\ref{sud15}),
which arises from the soft gluon interaction between the
initial and final states gauge links.
The comparison was made for single top (a) and single anti-top (b) quark production with two different $|y_J|$ regions.
It clearly shows that the color coherence effect
between the initial and final state colored particles
pushes $q_{\perp}$ to a smaller value.
Such effects become more important when the final state jet is required to be
in the forward region ($3\leq|y_J|\leq 4.5$).

Next, we examine the effect of the incoming bottom quark mass
to the $q_{\perp}$ distribution.
To be consistent with the (CT14) PDFs
which were determined using the S-ACOT scheme to define heavy parton,
Eq.~(\ref{eq:cf}) is used in the calculation of the $W$-piece. In the limit of
$m_b \to 0$, it reduces to the usual Wilson coefficient~\cite{Nadolsky:2002jr,Belyaev:2005bs,Berge:2005rv} (see Eq.~(\ref{eq:cf2})).
As shown in Fig.~\ref{fig:nb},  the correct treatment of the finite  bottom quark mass,
with $m_b=4.75$ GeV, shifts the peak of the $q_{\perp}$ distribution by about
$3\sim 4~{\rm GeV}$ as compared to massless case.

\begin{table*}[t]
\begin{center}
\caption{The predicted kinematic acceptances for the $\phi^*$ cut-off in the $t$-channel single top quark production at the LHC}
\begin{tabular}{c|c|c|c|c|c|c}
\hline
~~$\phi^* (t)$~~&~~$<0.05$~~&~~$<0.1$~~&~~$<0.15$~~&
~~$<0.2$~~&~~$<0.25$~~&~$<0.3$~\\
\hline
Res~~~~  $|y_J|<4.5$& 48\%& 68 \%& 79\%& 86\%& 91\%& 94\%\\
\hline
PYTHIA $|y_J|<4.5$& 46\% & 68\% & 81\% & 88\% &  93\%&  96\%  \\
\hline
\hline
Res~~~~  $3<|y_J|<4.5$& 54\%& 72\%& 83\%& 89\%& 93\%& 96\%\\
\hline
PYTHIA $3<|y_J|<4.5$& 46\% & 68\% & 80\% & 87\% &  92\%&  96\%  \\
\hline

\hline
~~$\phi^* (\bar{t})$~~&~~$<0.05$~~&~~$<0.1$~~&~~$<0.15$~~&
~~$<0.2$~~&~~$<0.25$~~&~$<0.3$~\\
\hline
Res~~~~  $|y_J|<4.5$& 48\%&  68\%& 79\%& 86\%& 90\%& 94 \%\\
\hline
PYTHIA $|y_J|<4.5$& 48\% & 70\% & 82\% & 89\% &  93\%&  96\%  \\
\hline
\hline
Res~~~~  $3<|y_J|<4.5$& 54\%& 72\%& 82\%& 88\%& 92\%& 95\%\\
\hline
PYTHIA $3<|y_J|<4.5$& 48\% & 69\% & 81\% & 88\% &  93\%&  96\%  \\
\hline
\end{tabular}
\label{tbl:phicut}
\end{center}
\end{table*}

It is also desirable to compare the rapidity distribution of the final state jet in
various calculations.
Figure~\ref{fig:yj} shows the jet rapidity distributions predicted by our
resummation calculation (blue solid  line),  NLO (red dashed  line), PYTHIA (green solid  line),
and  LO calculation with CT14LO PDF (black dotted line), at the $\sqrt{S}=13~{\rm TeV}$ LHC,
with $|y_t|<3$ and $P_{J\perp}>30~{\rm GeV}$.
For this comparison, both
the renormalization and factorization scales are fixed at $H_T$ in the fixed-order calculation,
while $\mu_{\rm {Res}}=H_T$ and $\mu_F=b_0/b_*$ in the resummation calculation.
Firstly, we note that the NLO QCD correction modifies
the shapes of the $y_J$ with respect to PYTHIA prediction
which is based on the leading order matrix element calculation.
The similar result was discussed in Ref.~\cite{Berger:2017zof}.
This is due to the sizeable NLO corrections originated  from gluon scattering sub-processes
which generate a large correction to the central jet rapidity  region and result in different 
shape between the NLO and LO results.
Secondly, the differential cross section predicted by the
resummation calculation is  about the same as the NLO prediction.

As discussed above, the coherence effect of gluon radiation in the initial and finals states
becomes large when the final state jet falls into more forward (or backward) direction, with
a larger absolute value of pseudorapidity. Furthermore, a different prediction in $q_\perp$
would lead to different prediction in the azimuthal angle between the final state jet and the
top quark moving directions measured in the laboratory frame. Both of them suggest that we
could use the well-known $\phi^*$ distribution, for describing the precision
Drell-Yan pair kinematical distributions~\cite{Banfi:2010cf}, to test the effect of multiple gluon
radiation in the t-channel single top (anti-) quark production.
The advantage of studying the $\phi^*$ distribution is that it only depends on the
moving directions (not energies) of the final state jet and top (anti-) quark.
Hence, it might provide a more sensitive experimental observable when the
final state jet falls into forward (or backward) direction.
We follow its usual definition and define
\bea
\phi^*=\tan\left(\frac{\pi-\Delta\phi}{2}\right) \sin \theta^*_\eta,
\eea
where $\Delta\phi$ is the azimuthal angle separation in radians
between the jet and top quark. The angle $\theta^*_\eta$ is defined as,
\bea
\cos\theta^*_\eta=\tanh\left[\frac{\eta_J-\eta_t}{2}\right],
\eea
where $\eta_J$ and $\eta_t$ are the pseudorapidities of the jet and top quark, respectively. Here, we used pseudorapidity, instead of rapidity, of top quark because rapidity depends on the energy of particle. 

As shown in Fig.~\ref{fig:phi}, the predictions of PYTHIA and our resumamtion calculation
differ in the small $\phi^*$ region, especially for the final state jet falls into more  forward (or backward) direction (Fig.~\ref{fig:phi}(b, d)),  which can be caused by a large value of $\eta_J-\eta_t$. i.e.,
in the events with large rapidity gap. In such region, the subleading logarithm terms in
the Sudakov factor are important in our resummation calculation. To illustrate this, we also compare to the
prediction (shown as black curves in  Fig.~\ref{fig:phi}) without the coherence factor $T$ in Eqs. (\ref{sud12})-(\ref{sud15}).
It shows that factor $T$ would change $\phi^*$ distribution significantly.

Since  $\phi^*$ is sensitive to the color structure of the signal, it could also be used to improve the $t$-channel single top quark cross section measurement. In  that case,  a precise theoretical evaluation of the kinematic acceptance after imposing the kinematic cuts is necessary, which is defined as,
\bea
\epsilon\equiv\dfrac{\sigma(\phi^*<\phi^0)}{\sigma}.
\eea
Here, $\sigma(\phi^*<\phi^0)$ is the cross section after imposing the kinematic cuts, while $\sigma$ is not.
As shown in Table~\ref{tbl:phicut},  if we require the final state jet to be in the forward rapidity region, with $3\leq|y_J|\leq 4.5$, the  kinematic acceptance with $\phi^*<0.05$ is larger by about 15\% for top quark and  11\% for anti-top quark in our resummation calculation than the PYTHIA prediction. For $\phi^*<0.1$, they differ by about 6\% for top quark and 4\% for anti-top quark,  and our resummation calculation predicts a larger total fiducial cross section. Currently, the ATLAS and CMS Collaborations have measured the $t$-channel single top quark  at the 13 TeV LHC,  the uncertainty is around 10\%~\cite{CMS:2016ayb,Aaboud:2016ymp}. If the $\phi^*$ observable is used to further suppress backgrounds and enhance the signal to backgrounds ratio, the difference found in our resummation and PYTHIA calculations of the fiducial cross section could become important.  This will lead to, for example, different conclusion about the constraints on various $Wtb$ anomalous couplings, induced by New Physics, or the measurement of $V_{tb}$~\cite{Cao:2015doa,Cao:2015qta}.

\section{Conclusions}
In this paper, we studied the $q_\perp$ resummation effects for the $t$-channel single top quark production at the LHC based on the TMD factorization theorem.  The  large logarithm $\ln(Q^2/q_{\perp}^2)$ was resummed by renormalization group evolution. In order to validate our resummation formula, we expand it to the NLO  to obtain the singular terms and compare the transverse momentum distributions at the NLO level. It shows perfect agreement in the small $q_{\perp}$ region. We also calculate the NLO  total cross section based on the resummation framework, and our results are in perfect  agreement with  MCFM.

We then perform the calculation of the $q_{\perp}$ distribution at NLL accuracy, and compare them with predictions from PYTHIA.  It shows that the Sudakov  peak in this process is sensitive to the soft gluon interaction between the initial and final states, and the bottom quark mass.
Furthermore, we find the shape of $q_\perp$ spectrum from the our resummation calculation  agrees well  with PYTHIA results
when the final state jet is allowed to fall into the full rapidity region, but there is a large deviation when the final state jet is required to be in the forward (or backward) region. The rapidity distribution of the final state jet in various calculation are also discussed. We note that the NLO QCD correction modifies the shapes of the $y_J$ with respect to the PYTHIA and LO predictions, and the resummation calculation presented in this work is at the NLO-NLL accuracy.

Finally, we propose to measure the  experimental observable $\phi^*$, similar the one used in analyzing the precision Drell-Yan data, to  perform precision test of the SM in the production of the $t$-channel single top events at the LHC. It shows the predictions of PYTHIA and our resummation calculation differ in the small $\phi^*$ region, especially when the final state jet falls into more forward (or backward) region. The difference found in our resummation and PYTHIA calculations of the fiducial cross section could become important if $\phi^*$ is used to further suppress the  backgrounds in order to determine the CKM mixing-matrix element $V_{tb}$ or to probe new physics effect through measuring the $Wtb$ couplings in the t-channel single-top events produced at hadron colliders.

\begin{acknowledgments}
This work is partially supported by the U.S. Department of Energy,
Office of Science, Office of Nuclear Physics, under contract number
DE-AC02-05CH11231; by the U.S. National
Science Foundation under Grant No. PHY-1719914; and
by the National Natural Science Foundation of China under Grant
Nos. 11275009, 11675002, 11635001 and 11725520.
C.-P. Yuan is also grateful for the support from 
the Wu-Ki Tung endowed chair in particle physics.
\end{acknowledgments}

\bibliographystyle{apsrev}
\bibliography{reference}

\end{document}